\documentclass[linenumber]{aastex631}
\usepackage[utf8]{inputenc}
\usepackage[version=3]{mhchem} 
\usepackage{graphicx} 
\usepackage{natbib} 
\usepackage{amsmath} 
\usepackage{graphicx}
\usepackage[outdir=./]{epstopdf}
\usepackage{natbib}
\usepackage{multirow}
\usepackage{xcolor}
\usepackage[flushleft]{threeparttable}
\usepackage{newtxtext,newtxmath}
\usepackage{graphicx,subfigure}
\usepackage{multirow}
\usepackage{tabulary,booktabs}
\usepackage{amsmath}
\usepackage{hyperref}
\usepackage{longtable}
\usepackage{xspace}
\usepackage{array}
\usepackage{epstopdf}
\usepackage{datetime2}

\newcommand{\tess}{{\em\rm TESS\/}}
\newcommand{\dsct}{\mbox{$\delta$~Sct}}
\newcommand{\kepler}{{\em\rm Kepler\/}}

\newcommand{\Gaia}{{\em\rm Gaia\/}}

\begin{document}

\title{Signature of high-amplitude pulsations in seven \dsct\ stars via \tess\ observations}

\author[0000-0001-8255-0871]{Fatemeh \textsc{Vasigh}}
\affiliation{Department of  Physics, Faculty of Science, University of Zanjan, University Blvd., Zanjan, 45371-38791, Zanjan, Iran}
\correspondingauthor{Elham \textsc{Ziaali}, Hossein \textsc{Safari}}
\email{ziaali@znu.ac.ir}, \email{safari@znu.ac.ir}

\author[0000-0001-7754-1452]{Elham \textsc{Ziaali}}
\affiliation{Department of  Physics, Faculty of Science, University of Zanjan, University Blvd., Zanjan, 45371-38791, Zanjan, Iran}
\affiliation{Instituto de Astrofísica de Andalucía -- CSIC, E-18008 Granada, Spain}%
\author[0000-0003-2326-3201]{Hossein \textsc{Safari}}
\affiliation{Department of  Physics, Faculty of Science, University of Zanjan, University Blvd., Zanjan, 45371-38791, Zanjan, Iran}
\affiliation{Observatory, Faculty of Science, University of Zanjan, University Blvd., Zanjan, 45371-38791, Zanjan, Iran}



\begin{abstract}
The regular behavior of the pulsations of high-amplitude \dsct\ (HADS) stars gives a greater chance to investigate the interiors of stars. We analyzed seven HADS stars showing peak-to-peak amplitudes of more than 0.3 mag that were newly observed by \tess. We obtained that TIC 374753270, TIC 710783, and TIC 187386415 pulsate in fundamental radial mode; also, TIC 130474019 and TIC 160120432 show double radial modes. On the other hand, TIC 148357344 and TIC 278119167 demonstrate triple-mode behavior. Our analysis shows that these seven stars are close to the red edge of the (inside) instability strip in the Hertzsprung–Russell diagram. 
The fundamental mode of these seven targets follows the period-luminosity (PL) relation for \dsct\ stars. However, TIC 278119167 deviates slightly from the fundamental PL relation. The double-mode and triple-mode HADS stars (TIC 130474019, TIC 160120432, TIC 148357344, and TIC 278119167) are in agreement with the period ratio ranges (fundamental to first and second overtones). 
Using the information of 176 HADS stars (Netzel and Smolec), we find a scaling relation (${\rm [Fe/H]} \propto \log (M^{7.95\pm 0.15} L^{-1.83\pm 0.11} {\rm P0^{0.79\pm 0.14}} T_{\rm eff}^{0.047\pm 0.02}$)) between the metallicity ([Fe/H]) and mass ($M$), luminosity ($L$), effective temperature ($T_{\rm eff}$), and the fundamental period ($\rm P0$). We estimate the metallicity of the seven newly identified HADS stars ranging from -0.62 to 0.37 dex.

\end{abstract}



\section{Introduction}\label{sec:intro}
Asteroseismology is an approach to studying the internal characteristics of pulsating stars through their oscillation modes \citep{aerts2010, Antoci2019, Tim2020, roAp2024}. The asteroseismology approach has more details explanations for the interior structure of different stars by increasing the resolution (temporal, spectral, and spatial) of ground and space observations of the pulsating stars. While overall instrument resolution (temporal and spatial) helps measure pulsation frequencies more precisely, the spectral resolution provides extra detail of chemical abundances (metallicity).
Studies of \kepler\ \citep{Gilliland2010} data have had significant impacts on asteroseismology. Recently, the Transiting Exoplanet Survey Satellite (\tess; \cite{Ricker2015}) provided unprecedented space observations of numerous star systems.
\cite{Antoci2019} reported the information of 117 \dsct\ stars from \tess\ short-cadence observations for sectors 1 and 2. Very recently, \citet{Amelie2024} identified 848 \dsct\ stars having a narrow color range of 0.29 $< G_{BP} - G_{RP} <$ 0.31 (at the center of instability strip) as being nearer than 500 pc by using sectors 27$\text{-}$55 of \tess\ observations and \Gaia\ Data Release 3 (DR3) information. Investigating the pulsation property of \dsct\ stars could help to understand their internal structure, such as rotation, chemical composition, density, temperature, and gravity.

In the Hertzsprung–Russell (H-R) diagram, \dsct\ stars are pulsating variables of spectral types A0$\text{-}$F5 that mainly lie at the overlap of the main sequence and the instability strip \citep{Breger_1979,murphy2020, Bowman2021}. 
The \dsct\ stars have the intermediate mass located at the transition region from low-mass to high-mass stars. Stars with lower than 1 and higher than 2 $M_{\odot}$ are called low-mass and high-mass, respectively. The effective temperature range of the \dsct\ stars lies between $6500$ and $9500$ K \citep{Petersen1996,Breger2000A, McNamara2011, Bowman2016, Pietrukowicz2020, murphy2019, jayasinghe2020, Yang2021b}.
The \dsct\ stars pulsation frequencies are in the range of 5-60\,$\rm day^{-1}$. They show various modes of pulsations such as radial, nonradial pressure modes, and mixed modes of the low radial order in their complex light curves \citep{Houdek.1999, Breger_2000B, Handler_2009,Hasanzadeh_2021, Soszynski2021}

Their pulsations are primarily driven by the classical $\kappa$ mechanism in the He~{\scriptsize \rm II} ionization zone. Some studies also suggested that intrinsically stable, stochastically driven (solar-like) $p$-modes may be excited simultaneously in such \dsct\ star \citep{Baker1962, Balmforth1990, Houdek.1999, samadi.2002, antoci2011, antoci2014}.

High-amplitude \dsct\ (HADS) stars pulsate in $V$-band amplitudes over 0.3 mag. However, there are few reports on HADS stars showing peak-to-peak value slightly smaller than 0.3 mag \citep{Breger2011, Lv_2023}. HADS stars are characterized by their high amplitude, short periods (less
than one day), and also $v\sin i \leq\ 30\ \rm km s^{-1}$. HADS stars oscillate as single-periodic with a radial fundamental mode or double-periodic or triple-periodic pulsators \citep{Bono1997,Wils2008, McNamara2011, Yang2021a, Netzel2022, Lv_2023, Xue2023}. However, the number of HADS stars having four independent radial frequencies as quadruple-mode HADS stars has been increasing in recent years \citep{Mow_2016, Lv2022}.

Also, nonradial modes may exist in the HADS frequency spectrum \citep{uytterhoeven.2011}.
HADS stars are important because they can be used to study the structure and evolution of stars. These stars can be used to investigate the evolution of stellar populations, as they are found in old stellar populations, while low-amplitude \dsct\ (LADS) stars are young and intermediate-age stars \citep{Chang2013,Netzel2022, Soszynski2021}.
Recently, there has been a growing interest in HADS stars as potential targets for exoplanet searches. High-precision observations of HADS stars' pulsations and signatures of orbiting planets could detect slight variations in their light curves. Several exoplanets have already been discovered around HADS stars \citep{Hey2021, Guzik2021}.
\citet{Lv_2023} studied seven new HADS stars through photometric data of \tess and reported that two stars may be RR Lyrae based on their light curves and the difference in their placement with the HADS star in the Peterson and period-luminosity (PL) diagrams. In these HADS stars, the metallicity can have a greater effect than the rotation of the star on the period ratios (the ratio of overtones to the fundamental mode). The period ratios decrease with increasing metallicity.

Metallicity is often expressed as a ratio of the abundance of iron (Fe) to hydrogen (H) compared to the solar ratio, written as [Fe/H] \citep{Kotoneva2002, Chruslinska2019}. A metallicity value greater than -0.5 dex indicates metal-rich \dsct\ stars, while a value less than -1.5 dex shows a metal-poor star \citep{McNamara2011}. Determining a star's metallicity involves sophisticated astronomical techniques based on spectroscopic investigations \citep{Lianou2011, Liu2020}, photometric methods \citep{Li_2023, Istvan2022} and astroseismology approach \citep{Netzel2022}. 

Here, we analyzed the light curves of seven newly identified HADS stars from \tess\ observations. We selected these HADS from the \tess\ catalog of several thousand of \dsct\ stars applying three criteria: (1) the peak-to-peak amplitude is greater than 0.3 mag, (2) the related extinction is less than 0.6 mag, and (3) the relative parallax error is less than 0.05.
Therefore, we applied more analyses to seven HADS having short-cadence \tess\ observation that satisfied the above criteria for more asteroseismic analyses. To do this, we extracted the frequencies and amplitudes of these seven stars using the periodogram analysis. 

We studied the PL behavior and H-R diagram. To determine the metallicity of targets, we obtained a scaling relation between the metallicity of 176 HADS stars \citep{Netzel2022} and their physical parameters (mass, luminosity, and effective temperature) and the fundamental periods.
Section \ref{sec:Data} gives the HADS star information used in this paper. Sections \ref{sec:Method} and \ref{sec:Results} explain the methods and results, respectively. Finally, section \ref{sec:conclusion} summarizes the essential findings of the present work.

\section{Data} \label{sec:Data}
\tess\ is a space mission for probing exoplanets (\cite{Ricker2015}). This telescope divides the sky into sectors and observes each sector for 27 days. The \tess\ space telescope was designed to record objects in a red passband of approximately 600–1000 nm \citep{Silva_Aguirre_2015, Barclay2018}.
\tess\ data are available in two forms, target pixel file (TPF) and light curve, which can be searched applying the Lightkurve tool \citep{Lightkurve}. Figure \ref{Figure 1} illustrates TPF and aperture (red squares) for TIC 374753270 (left, first row), TIC 130474019 (right, first row), TIC 148357344 (left, second row), TIC 160120432 (right, second row), TIC 278119167 (left, third row), TIC 710783 (right, third row), and TIC 187386415 (fourth row). As shown in the figure, each aperture covers the primary pixels of each target without considerable contamination of another object in the field of view.
In this paper, we used photometric data processed by SPOC \citep{Jenkins2016}, which are short-cadence (2 minutes). We transformed the flux to magnitude and subtracted from its average value for each light curve to get the corrected time series.

\begin{table}
	\centering
          \caption {TIC number, variable name, and \tess\
 short-cadence sector(s) for seven HADS stars.\label{tab: Table 1}}

	\begin{tabular}{|c c c|} 
		\hline
	TIC & Star Name & Sector\\
	\hline
	374753270&-&36,37\\
	130474019&-&7, 34\\ 
        148357344&-&7\\ 
	160120432&YZ UMi&14, 19, 20, 40, 47, 52, 53\\ 
        278119167&-&40, 53\\
        710783&-&5\\
        187386415&-&51\\
	\hline
	\end{tabular}
\end{table}

    http://dx.doi.org/10.17909/4a25-pg87

\section{Method}\label{sec:Method}
\subsection{Frequency analysis}\label{subsec: Frequency analysis}
We selected seven \dsct\ light curves (related to TPFs of Figure \ref{Figure 1}) with peak-to-peak amplitudes greater than 0.3 mag as an essential characteristic for identifying HADS stars. We checked the TPFs for each target to ensure the SPOC light curve's validity due to the SPOC aperture's size, which would not be contaminated by the neighbor objects in the field. Our investigation showed that the SPOC light curves were perfect for our targets. Then, their SPOC-prepared light curves were analyzed to extract pulsation frequencies. We removed outliers from each star's light curve. We rejected the outlier data points from light curves (data points were far from the average). We first calculated the average and standard deviation for each light curve. The outlier data points have amplitudes more significant than 4 standard deviations (4$\sigma$).
Also, the mean of each single light curve was subtracted to get the corrected systematic error flux.
To study our samples' pulsating behavior, the PERIOD04 \citep{Lenz_Breger2005} software was applied. A prewhitening procedure carried out the frequency extraction.
The Nyquist frequency is half of the sampling rate, $f_N$ = 360.048 $\rm day^{-1}$ for 2 minute cadence observations. This value is well above the limit range (5 $< f <$ 60 $\rm day^{-1}$) expected for a typical \dsct\ star pulsation.
We used a resolution frequency $f_{res}$ = 1.5/$\Delta$$T$ to identify nearby frequencies in the amplitude spectrum, where $\Delta$$T$ represents the length of the light curve \citep{Loumos_Deeming1978}. Two frequencies are resolved if the frequency difference is more than the resolution frequency.
The light curves are modeled by 
\begin{equation}\label{eq.1}
    m =m_0 + \Sigma_{i=1}^{N}\ A_i \sin(2\pi(f_it+\phi_i)),
\end{equation}

where $m_0$ is the zero point, $A_i$ is the amplitude, $f_i$ is the frequency, and $\phi_i$ is the phase. 

The PERIOD04 algorithm gives the frequency, amplitude, and phase of each significant mode by applying the multifrequency least-square fit (Equation \ref{eq.1}) for the stellar light curve. We considered the modes with a signal-to-noise ratio (S/N) more significant than 5.2 \citep{Baran2015}. The residual is determined by subtracting the modeled light curve from the original light curve that is expected to show the lack of significant modes (S/N$> 5.2$).
The uncertainty of the peaks is determined using the approach by \cite{Montgomery_O'Donoghuen1999}.

\begin{figure*}
     \centering
      \begin{subfigure}
         \centering
         \hspace*{0.01cm}
     \end{subfigure}\\
     \begin{subfigure}
         \centering
         \hspace*{0.01cm}
      \end{subfigure}\\
     \begin{subfigure}
         \centering
         \hspace*{0.01cm}
     \end{subfigure}\\
    \begin{subfigure}
       \centering
    \end{subfigure}
\end{figure*}
\begin{figure}
    \centering
    \includegraphics[width=.795\textwidth]{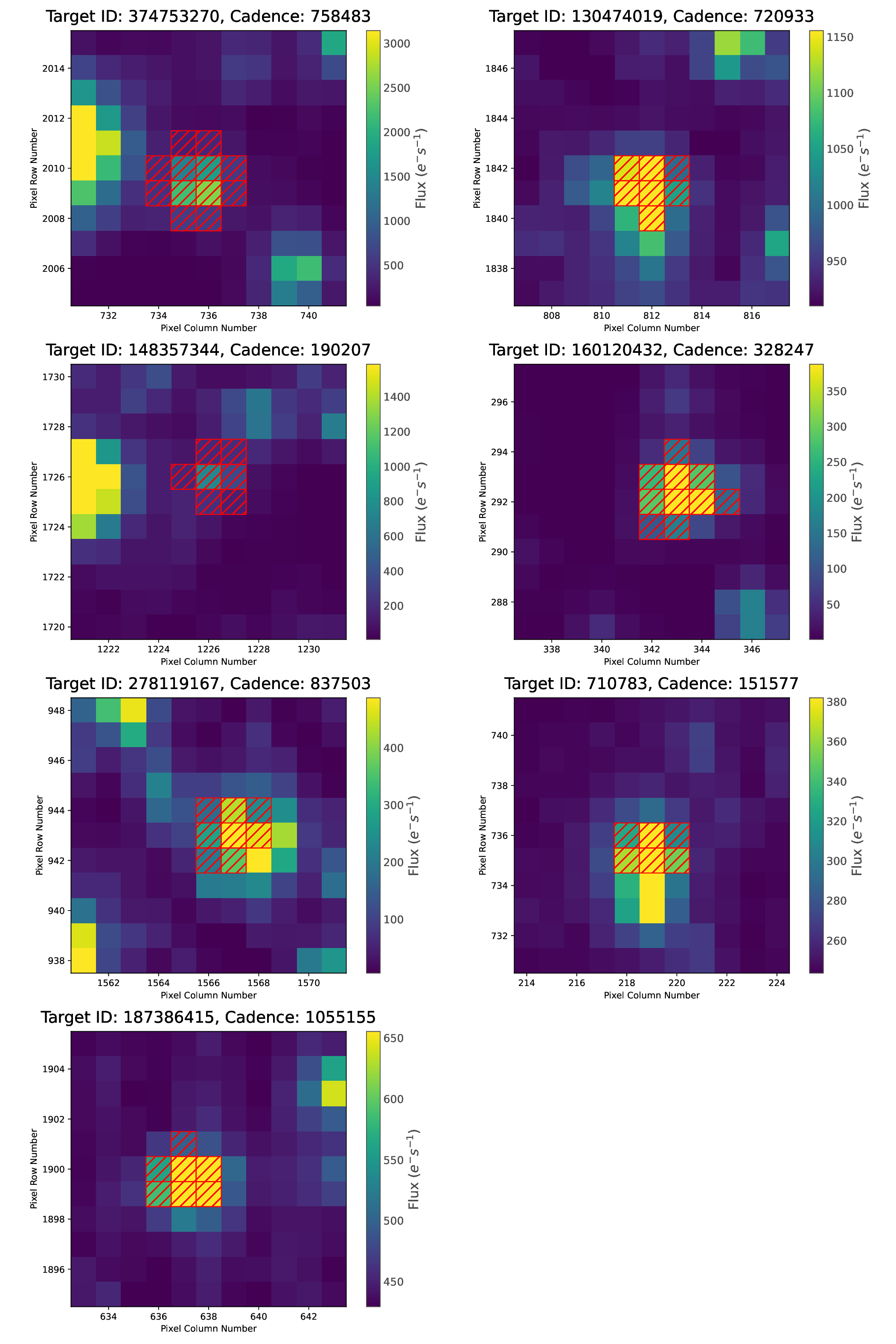}
    \caption{Target Pixel File (TPF) for TIC 374753270 (left-first row), TIC 130474019 (right-first row), TIC 148357344 (left-second row), TIC 160120432 (right-second row), TIC 278119167 (left-third row), TIC 710783 (right-third row), and TIC 187386415 (fourth-row). The aperture for each target star is indicated by the red squares.}
    \label{Figure 1}
\end{figure}

\subsection{Asteroseismic analysis}\label{subsec: Asteroseismic analysis}
The stellar oscillations are helpful to infer information about the stars' interior. Each mode can be described by three quantum numbers: $n$ (radial number), $l$ (degree), and $m$ (azimuthal number), which are related to the characteristics of oscillations.
In most pulsating stars, we expect the highest amplitude mode to be the fundamental radial ($n = 1$, $l = 0$) oscillation \citep{aerts2010, murphy2020}.

The \dsct\ stars' power spectrum, the frequency corresponding to the highest peak amplitude, is called $f_{A_{max}}$.
The $f_{A_{max}}$ might change over time due to modulation of amplitude and other factors such as noise \citep{Bowman2017}. 
Radial modes can be identified based on the ratio of their periods.
Frequency (period) ratios for radial modes (fundamental (F0), first (F1), second (F2) and third (F3) overtones) were first presented by \cite{Stellingwerf1979} based on theoretical models as $\rm F0/F1 = 0.756- 0.787$, $\rm F0/F2 = 0.611- 0.632$, and $\rm F0/F3 = 0.50- 0.525$.

For \dsct\ stars with the first and the second overtones, the characteristic frequency (period) ratios are around $\rm F1/F2 = 0.8$ \citep{Soszynski2021}.
The combination of different frequencies or the linear sum of the independent frequencies, $n\nu_{i} \pm m\nu_{j}$, was identified by the \cite{Loumos_Deeming1978} criteria. The modes with no agreement with combination or harmonic frequencies might be identified as nonradial modes. 
Whereas the frequency (period) of pulsations is linked to stars' interior characteristics, the frequency is related to the radius changes. The critical factor is the density of stars that is accurately related to the period. The period-density relation is defines as
\begin{equation}\label{eq.2}
   Q = P \sqrt{\overline{\rho}/ \overline{\rho}_{\odot}},
\end{equation}
where $\rm\overline{\rho}$ describes the mean densities of the target star and $\rm\overline{\rho}_{\odot}$ is the mean density for Sun. $P$ is the pulsation period.
The other forms of Equation (\ref{eq.2}) were given by \cite{Breger1990} that included the gravity acceleration and bolometric magnitudes.
Here, we used Equation (\ref{eq.2}) to validate the periods of seven targets. The range of pulsations constant ($Q$) for \dsct\ stars' fundamental, first, second, and third overtone frequencies are $0.027- 0.04$, $0.021-0.027$, $0.018- 0.021$, and $0.016- 0.017$, respectively \citep{Breger1975, Milligan_Carson1992}.

\subsubsection{Period–Luminosity relation}\label{subsec: pl}
The PL relation is a dependency between the star's luminosity and density. Consequently, this relation occurs between the luminosity and the pulsation period of a group of pulsating stars that have a relatively narrow range of effective temperature. The first PL was given by \citet{Leavitt1912} for Cepheid stars.
We computed the absolute $V$-band magnitudes ($M_{\rm V}$) for our seven targets using the following equation:
\begin{equation}\label{eq.3}
M_{\rm V} = m_{\rm v} + 5 \log \pi + 5 - A_{\rm V},
\end{equation}
where $m_{\rm v}$ is the apparent magnitude in the $V$ band and $\pi$ denotes the star's parallax (in arcseconds) collected from \Gaia\ DR3. Since interstellar dust can impact the flux of stars (luminosity), the $M_{\rm V}$ values were corrected for extinction ($A_{\rm V}$). Following \cite{Green2018}, we determined the extinction for our seven targets to correct the $V$-band magnitudes. All seven samples have absolute relative parallax error (the ratio of parallax uncertainty to the parallax) in the range of 0.010-0.020 and extinctions less than 0.6 mag.
By applying the Tycho $V_{T}$ and $B_{T}$ magnitudes, we calculated the Johnson $V$ apparent magnitudes as  \citep{hog2000, 1997ESA}, 
\begin{equation}\label{eq.4}
V = V_{T} - 0.090(B_{T} - V_{T}).
\end{equation}

We investigated the validity of the obtained fundamental frequencies by using the PL relation for our targets.
Using \Gaia\ Data Release 2 parallaxes, \citet{Ziaali_2019} gave the PL relation for \dsct\ stars observed by \kepler, as follows:
\begin{equation}\label{eq.5}
M_{\rm V}  = (-2.94 \pm 0.06) \log (P) + (-1.34 \pm 0.06),   
\end{equation}
While \citet{barak2022} reported a similar relation for stars observed by \tess\ using the \Gaia\ DR3 parallaxes,
\begin{equation}\label{eq.6}
M_{\rm V} = (-3.01 \pm 0.07) \log (P) + (-1.40 \pm 0.07). 
\end{equation}
Also, using multicolour photometric ground-based observations, \citet{Poro_2021} gave
\begin{equation}\label{eq.7}
M_{\rm V} = (-3.200 \pm 0.010) \log (P) -(1.599 \pm 0.010).
\end{equation}
By using Equation (\ref{eq.3}) the absolute magnitude for each single star was calculated, and the PL diagram is examined (Section \ref{subsec:Asteroseismical parameters}). These PL relations, Equations \ref{eq.5}-\ref{eq.7}, are very close together in a narrow ridge for \dsct\ stars. So, the magnitudes and fundamental periods should be consistent with the PLs within the observational uncertainties.

\section{Results}\label{sec:Results}
We applied three criteria to identify the HADS stars, including the peak-to-peak amplitude more significant than 0.3 mag (from the \tess\ catalog), relative parallax error less than 0.05 (from \Gaia\ DR3 catalog), and its measured extinction less than 0.6 mag.
Applying the frequency analysis for the \tess\ light curves, we investigated the properties of the amplitude spectrum. We considered the frequencies with S/N $>$ 5.2 to ensure the validity of the identified frequencies. We obtained the period ratio and {$Q$} value.
As essential criteria, if the targets (Table \ref{tab: Table 1}) have the \dsct\ star characteristic, we expect their absolute magnitudes and fundamental frequencies to satisfy the PL relations, Equations \ref{eq.5}-\ref{eq.7}. Please note that these PL relations are in good agreement with the slight differences that are derived from different observations.

\begin{figure}
    \centering
    \includegraphics[width=12.3cm,height=10cm]{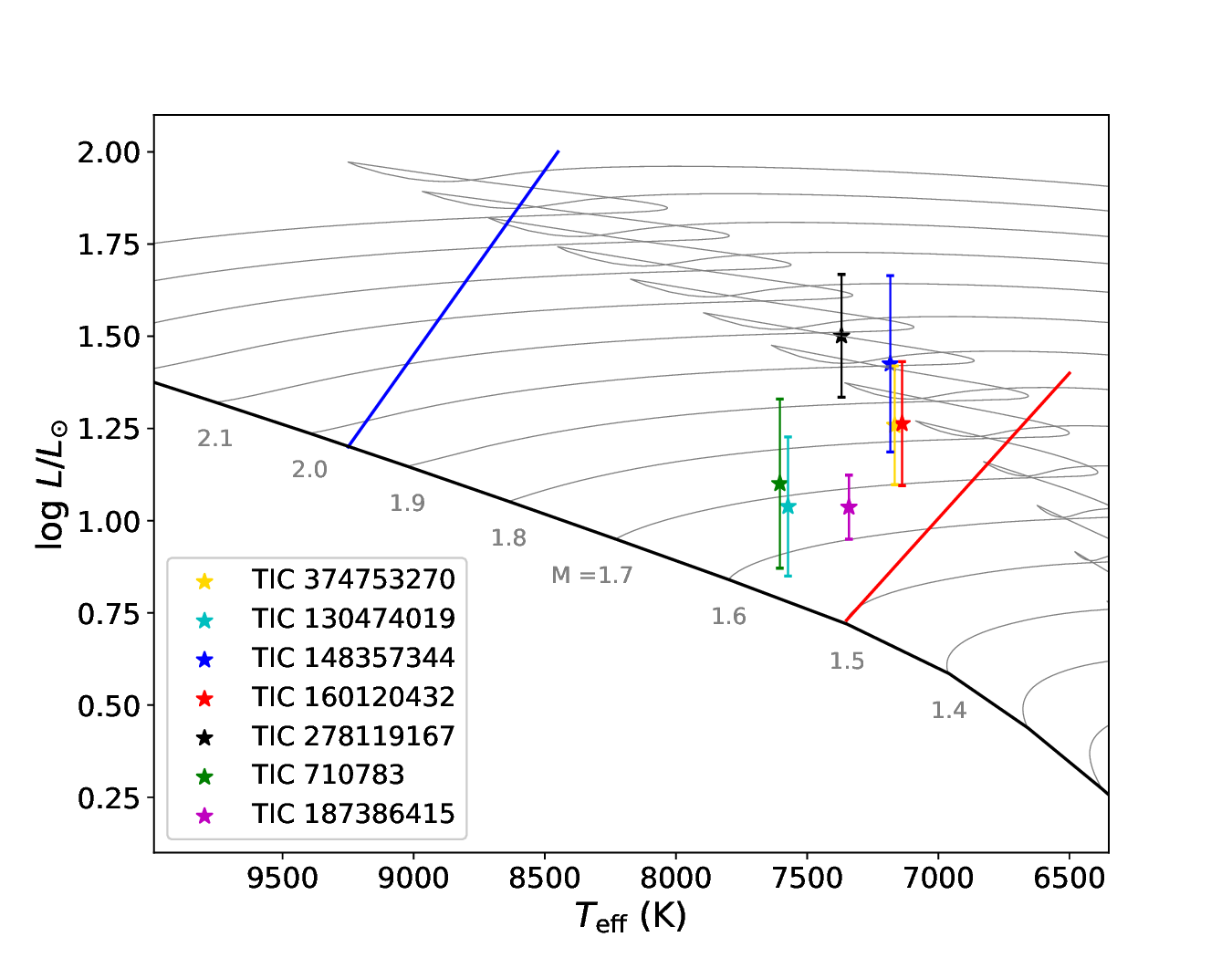}\hfill
    \caption{H-R diagram including our seven HADS stars indicated with different-colored star markers. The observational red and blue edge of the instability strip for \dsct\ stars along with the evolutionary tracks with $X=0.71$, $Z=0.014$, and $\alpha_{MLT}$=1.8 were obtained by \cite{murphy2019}. The temperatures are from \citet{Stassun2018} and the $V$ band luminosities were calculated by using the apparent magnitudes, parallaxes, and the interstellar extinctions (Eq. \ref{eq.3}). The error bars indicate the errors for luminosities.}\label{Figure 2}
\end{figure}

Figure \ref{Figure 2} displays the position of the seven HADS stars (colored star markers) between the instability strip edges in the H-R diagram given by \citet{murphy2019}. As displayed in the figure, these seven targets tend toward the instability strip's red edge with lower temperatures. \citet{murphy2019} investigated the instability strip boundaries based on the position of 15000 \dsct\ stars in the  \kepler\ field of view using time-dependent convection and mixing length ($\alpha_{MLT}$=1.8) reported by \cite{Dupret_2005}. In the remainder of this paper, we examined the HADS star criteria for these seven targets.


\subsection{TIC 374753270}
\begin{figure}
\centering
    \includegraphics[width=.7\textwidth]{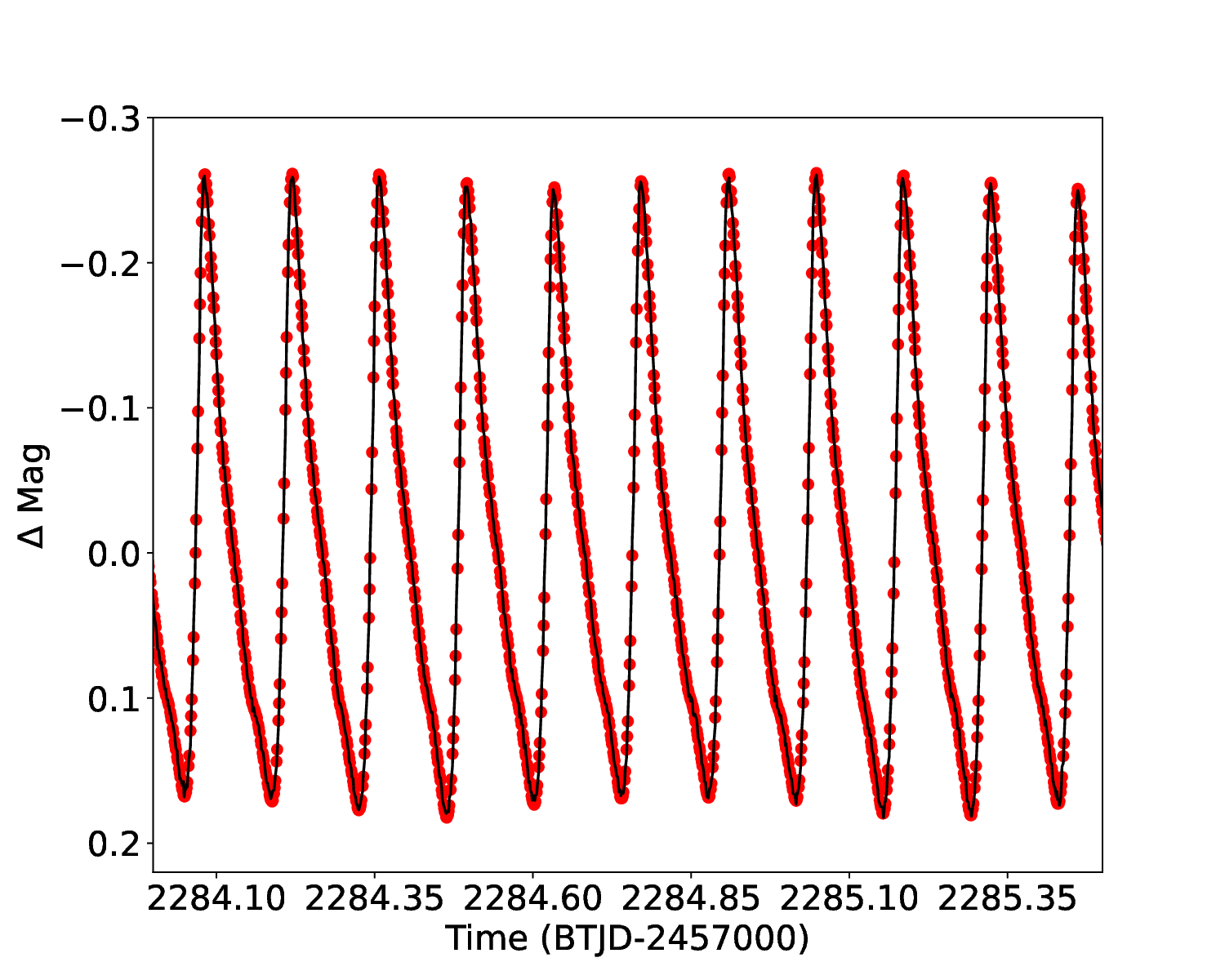}\hfill
    \\[\smallskipamount]
    \includegraphics[width=.7\textwidth]{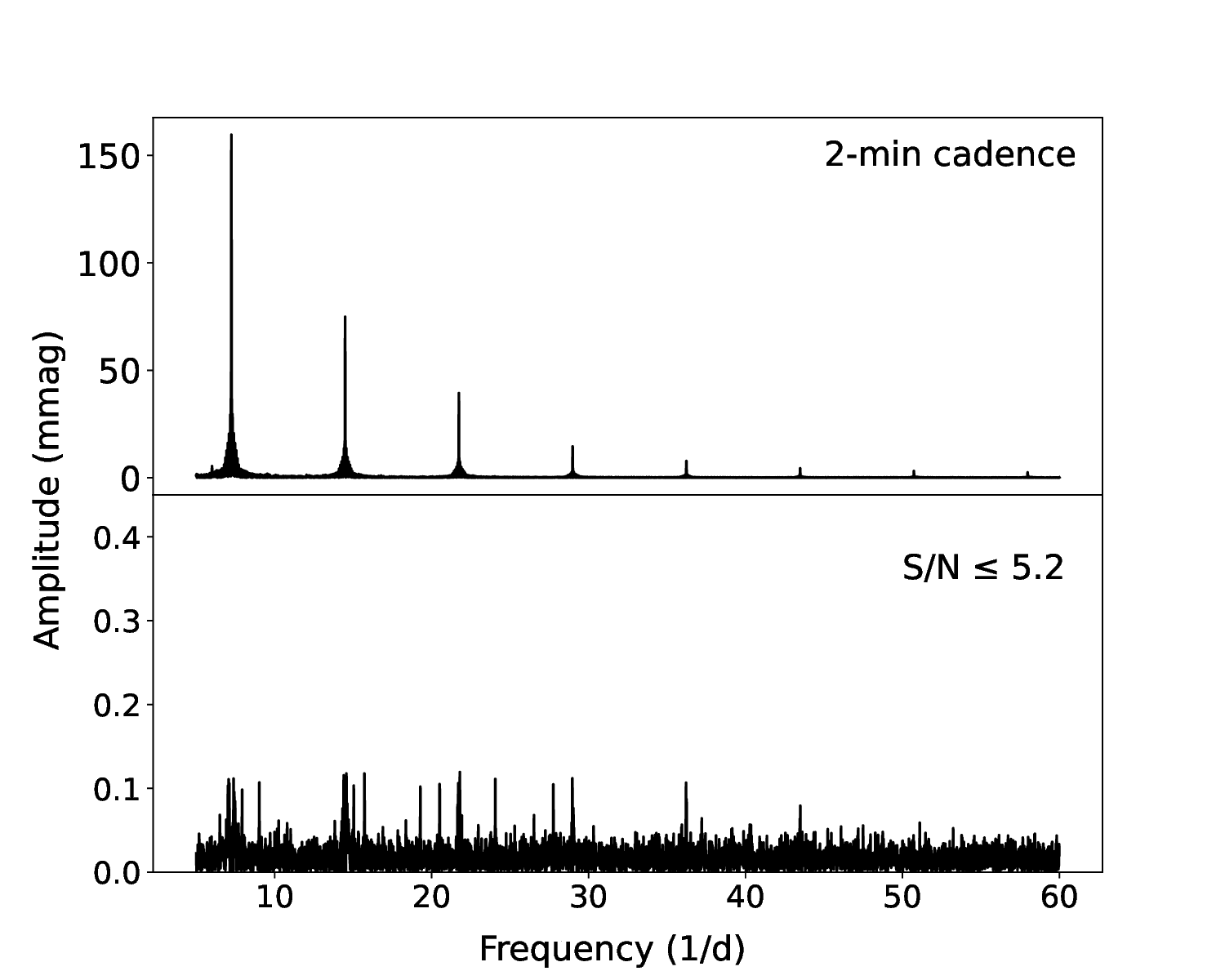}\hfill
    \caption{The short-cadence light curve (sectors 36 and 37) of TIC 374753270 during 1.5 days (top panel), amplitude spectrum (middle panel), and amplitude spectrum of residual with S/N less than 5.2 (bottom panel).}\label{Figure 3}
\end{figure}

Figure \ref{Figure 3} represents the light curve (top panel), amplitude spectrum (middle panel), and amplitude spectrum of the residual (bottom panel) for TIC 374753270 (TYC 8943-3384-1). \citet{barak2022} introduced TIC 374753270 ($R.A. = 10^h03^m20^s.62$, $DEC = -61^{\circ}51^{\prime}15.63^{\prime\prime}$) as a \dsct\ star. 
We obtained the significant frequencies for combined sectors 36 and 37 using the frequency analysis approach. By frequency analysis of the combined light curve, we detected 14 significant frequencies with S/N $> 5.2$ and frequency resolution greater than 0.029 $\rm day^{-1}$.
The frequencies are listed in the Table \ref{tab: Table 2}, including one radial frequency (monomode HADS) of $\rm F0 = 7.246052$ $\rm day^{-1}$ ($\rm P0$ = $0.138$ days). Other frequencies might be called harmonics ($f_{2}$, $f_{3}$, $f_{4}$, $f_{5}$, $f_{7}$, $f_{8}$, $f_{9}$) and nonradial modes ($f_{6}$, $f_{10}$, ..., $f_{14}$). These frequencies are in the reported range for \dsct\ stars \citep{Breger2000A,Liakos2017}.
Removing these 14 identified modes, we obtained the amplitude spectrum of the residual that indicated the absence of significant modes with S/N greater than 5.2.
The {$Q$} value of the radial fundamental mode is 0.032, which is in the reported range of fundamental mode for \dsct\ stars. 
The rectified light curve shows a peak-to-peak amplitude of about $\sim0.43$ mag, which is slightly more than 0.3 mag (expected for HADS stars).

\begin{table}
    	\centering
         \caption{A complete list of the 14 identified frequencies for TIC 374753270 (denoted by $f_{i}$). }\label{tab: Table 2}\
         
          \begin{tabular}{|c c c c c c c c|}  
		\hline
	$f_{i}$ & Frequency  & Amplitude  & phase  & S/N & ID & period ratio & {$Q$} value\\
		\hline
        
	 & ($\rm day^{-1}$) &  (mmag) &  (radians/2$\pi$) &  &  &  & \\
		\hline
	1 & 7.246052±0.000002 & 159.91±0.03 & 0.50029±0.00003 & 1430.94& F0 & -& 0.032±0.004\\
	2 & 14.491993±0.000004 & 74.91±0.03 & 0.31175±0.00006 & 1196.53& 2F0 & -&-\\
	3 & 21.737972±0.000008  & 39.41±0.03 & 0.9745±0.0001 & 825.07& 3F0 & - &-\\
        4 & 28.98394±0.00002 & 14.57±0.03 & 0.6471±0.0003& 394.95& 4F0 & - &-\\
        5 & 36.22991±0.00004 & 7.83±0.03 & 0.3280±0.0006 & 236.33& 5F0 & - &-\\
        6 & 6.01456±0.00005 & 5.81±0.03 & 0.6383±0.0008 & 142.08& nonradial&-  &- \\
        7 & 43.47588±0.00007 & 4.46±0.03 & 0.961±0.001 & 140.41& 6F0 & - &-\\
        8 & 50.7219±0.0001 & 3.14±0.03 & 0.264±0.001 & 101.92& 7F0  & - &-\\
        9 & 57.9679±0.0001 & 2.45±0.03 & 0.023±0.002 & 40.28&8F0 & - &-\\
        10 & 12.0289±0.0002 & 1.34±0.03 & 0.939±0.003 & 18.88& nonradial & - &-\\
	11 & 18.0437±0.0006 & 0.51±0.03 & 0.367±0.01 & 17.45&nonradial&-  &-\\
	12 & 13.2607±0.0007  & 0.42±0.03 & 0.059±0.01 & 15.06& nonradial &-  &-\\
        13 & 8.477±0.001 & 0.21±0.03& 0.324±0.02 & 6.42& nonradial&- &-\\
        14 & 20.506±0.001 & 0.18±0.03& 0.902±0.02 & 5.33& nonradial&- &-\\
		\hline
  \multicolumn{2}{l}{Note. The frequency resolution is about $f_{res}=0.029 \rm day^{-1}$.}
	\end{tabular}
\end{table}

\subsection{TIC 130474019}
\begin{figure}
\centering
    \includegraphics[width=.7\textwidth]{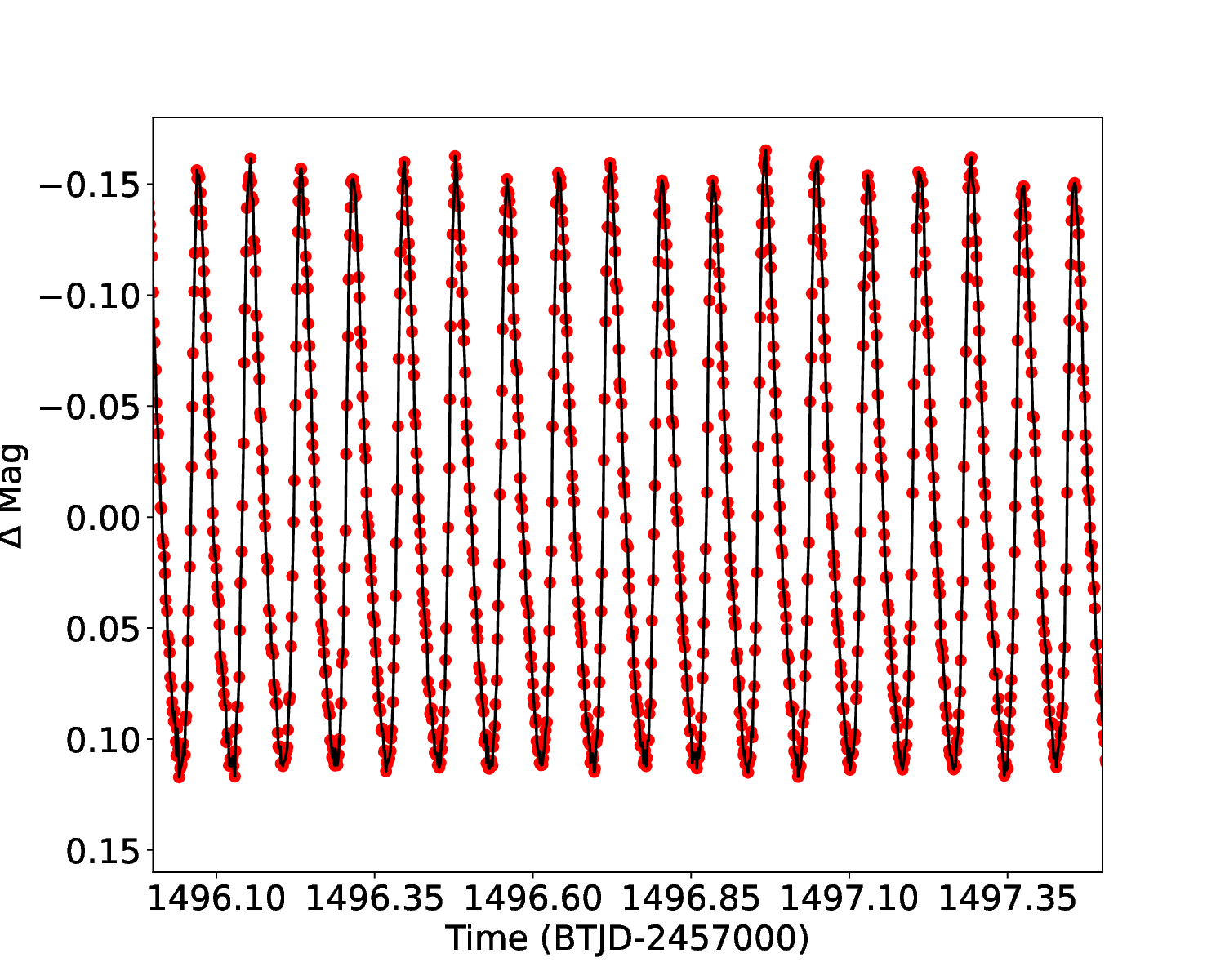}\hfill
    \\[\smallskipamount]
    \includegraphics[width=.7\textwidth]{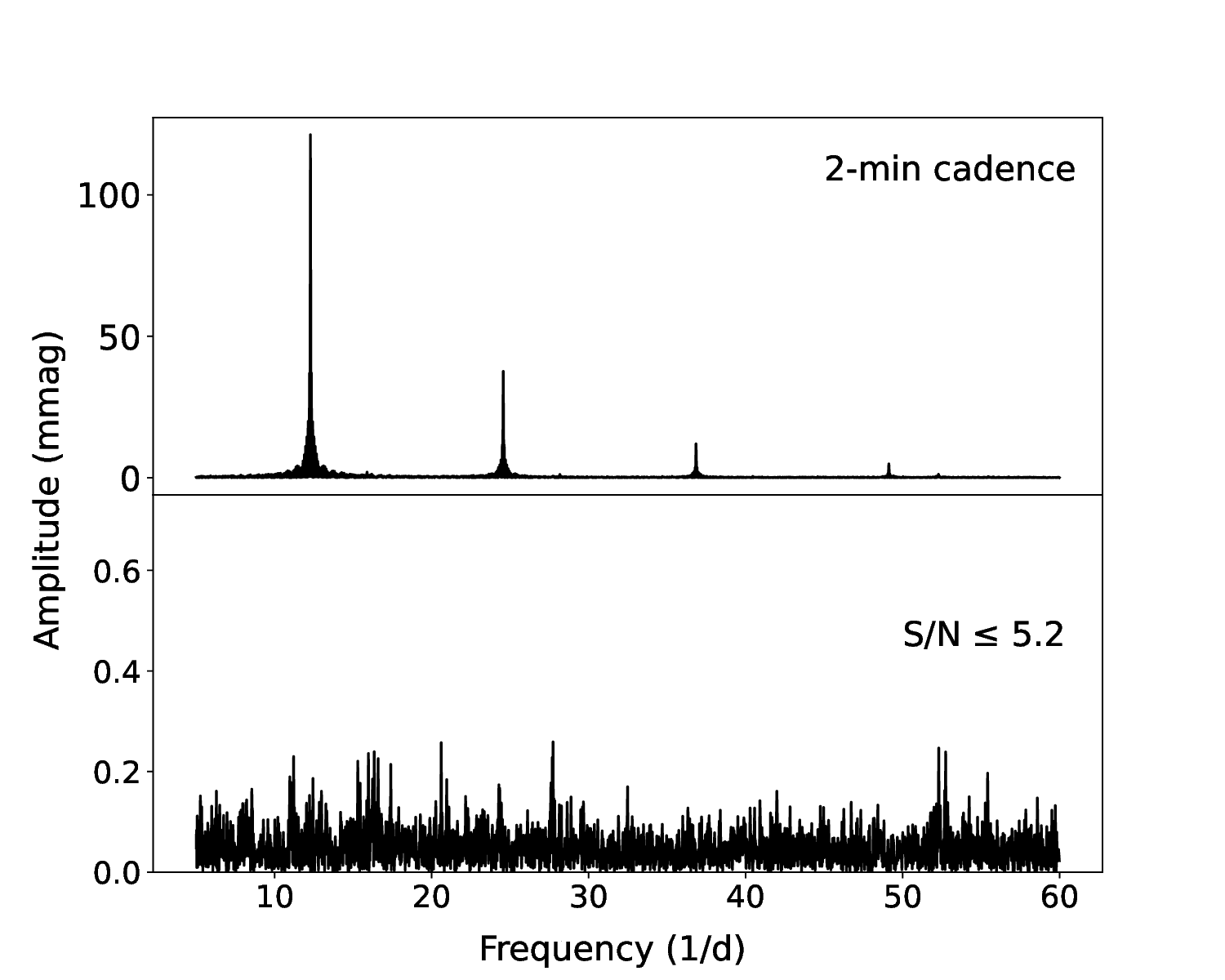}\hfill
   \caption{The short-cadence light curve (sector 7) of TIC 130474019 during 1.5 days (top panel), amplitude spectrum (middle panel), and amplitude spectrum of residual with S/N less than 5.2 (bottom panel).}\label{Figure 4}
\end{figure}

Figure \ref{Figure 4} shows the light curve (top panel), amplitude spectrum (middle panel), and amplitude spectrum of the residual (bottom panel) for TIC 130474019 (TYC 7636-1969-1). \cite{2020_barcelo} determined the physical parameters (temperature, gravity, and frequency scaling relation) of TIC 130474019 ($R.A. = +6^h59^m54^s.82$, $DEC = -42^{\circ} 01^{\prime} 15.12^{\prime\prime}$) using \tess\ observation, while \cite{barak2022} introduced this star as a \dsct\ sample. This star's light curves recorded in sectors 7 and 34 include 16339 and 16850 data points, respectively.
Table \ref{tab: Table 3} shows the 16 frequencies, which include two radial frequencies of $\rm F0 = 12.28231$ $\rm day^{-1}$ (fundamental, $\rm P0$ = $0.0814$ days) and $\rm F1 = 15.8872$ $\rm day^{-1}$ (first overtone) with {$Q$} values as 0.04 and 0.031 respectively. 
The frequency ratio ($\rm F0/\rm F1$) is obtained to be 0.78, which indicates the independent property and validity of the fundamental and first overtone.
As tabulated in Table\ref{tab: Table 3}, we observed the harmonics frequencies ($f_{2}$, $f_{3}$, $f_{4}$), combination frequencies ($f_{6}$, $f_{10}$, $f_{11}$, $f_{13}$), and nonradial modes ($f_{7}$,.., $f_{9}$, $f_{12}$, $f_{14}$,..., $f_{16}$).
The peak-to-peak amplitude is about $\sim 0.3$ mag, and the range of the frequencies might suggest TIC 130474019 is a double-mode HADS star.

\begin{table}
    \centering
    \caption{A complete list of the 16 identified frequencies for TIC 130474019 (denoted by $f_{i}$). }\label{tab: Table 3}\
    \begin{tabular}{|c c c c c c c c|}  
	\hline
	$f_{i}$ & Frequency & Amplitude  & phase & S/N & ID & period ratio & {$Q$} value\\
		\hline
   & ($\rm day^{-1}$) &  (mmag) &  (radians/2$\pi$)& &  &  & \\
  \hline
	1 & 12.28231±0.00001 & 121.25±0.06 & 0.95194±0.00008& 1372.4 & F0 &- & 0.04±0.004\\
	2 & 24.56457±0.00003 & 37.73±0.06 & 0.9304±0.0002& 639.01 & 2F0 & - &-\\
	3 & 36.8468±0.0001 & 11.96±0.06 & 0.7588±0.0008& 279.6 & 3F0 &- &-\\
        4 & 49.1293±0.0002 & 4.92±0.06 & 0.375±0.002& 106.2 & 4F0 & -&-\\
        5 & 15.8872±0.0007 & 1.94±0.06 & 0.499±0.005& 18.10 & F1 & 0.78 & 0.027±0.003\\
        6 & 28.173±0.001 & 1.17±0.06 & 0.904±0.008& 12.57 & F0+F1 & - &-\\
        7&16.577±0.002&0.69±0.06&0.70±0.03&8.31& nonradial &-&-\\
        8 & 16.015±0.002 & 0.60±0.06 & 0.44±0.01& 7.96 & nonradial & - &-\\
        9 & 15.416±0.002 & 0.63±0.06 & 0.23±0.01& 8.29 & nonradial &- &- \\
	10 & 8.672±0.002 & 0.47±0.06 & 0.02±0.02& 7.18 &2F0-F1 &- &-\\
	11 & 40.452±0.002 & 0.46±0.06 & 0.71±0.02& 9.34 & 2F0+F1 & -&-\\
        12 &28.299±0.003 & 0.41±0.06 & 0.99±0.02& 8.24 & nonradial & - &- \\
        13 & 52.737±0.003&0.34±0.06&0.09±0.02&8.31&3F0+F1&-&-\\
        14&15.973±0.003&0.25±0.06&0.90±0.03&5.67&nonradial&-&-\\
        15&20.611±0.003&0.23±0.06&0.62±0.03&5.44&nonradial&-&-\\
        16&27.699±0.003&0.22±0.06&0.43±0.03&5.23&nonradial&-&-\\
        
		\hline
  \multicolumn{2}{l}{ Note. The frequency resolution is about $f_{res}=0.061 \rm day^{-1}$.}
	\end{tabular}
\end{table} 

\subsection{TIC 148357344}
\begin{figure}
\centering
     \includegraphics[width=.7\textwidth]{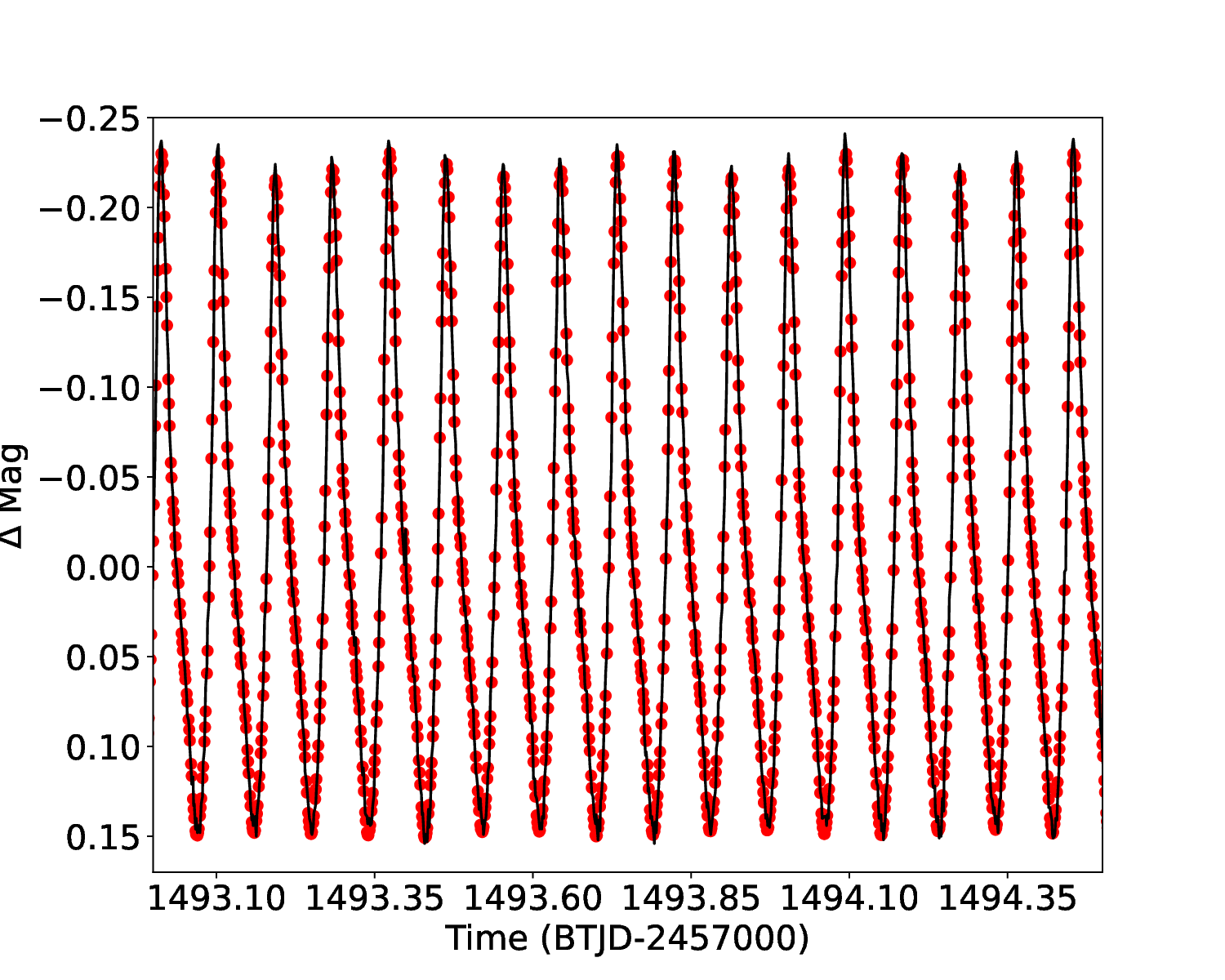}\hfill
    \\[\smallskipamount]
    \includegraphics[width=.7\textwidth]{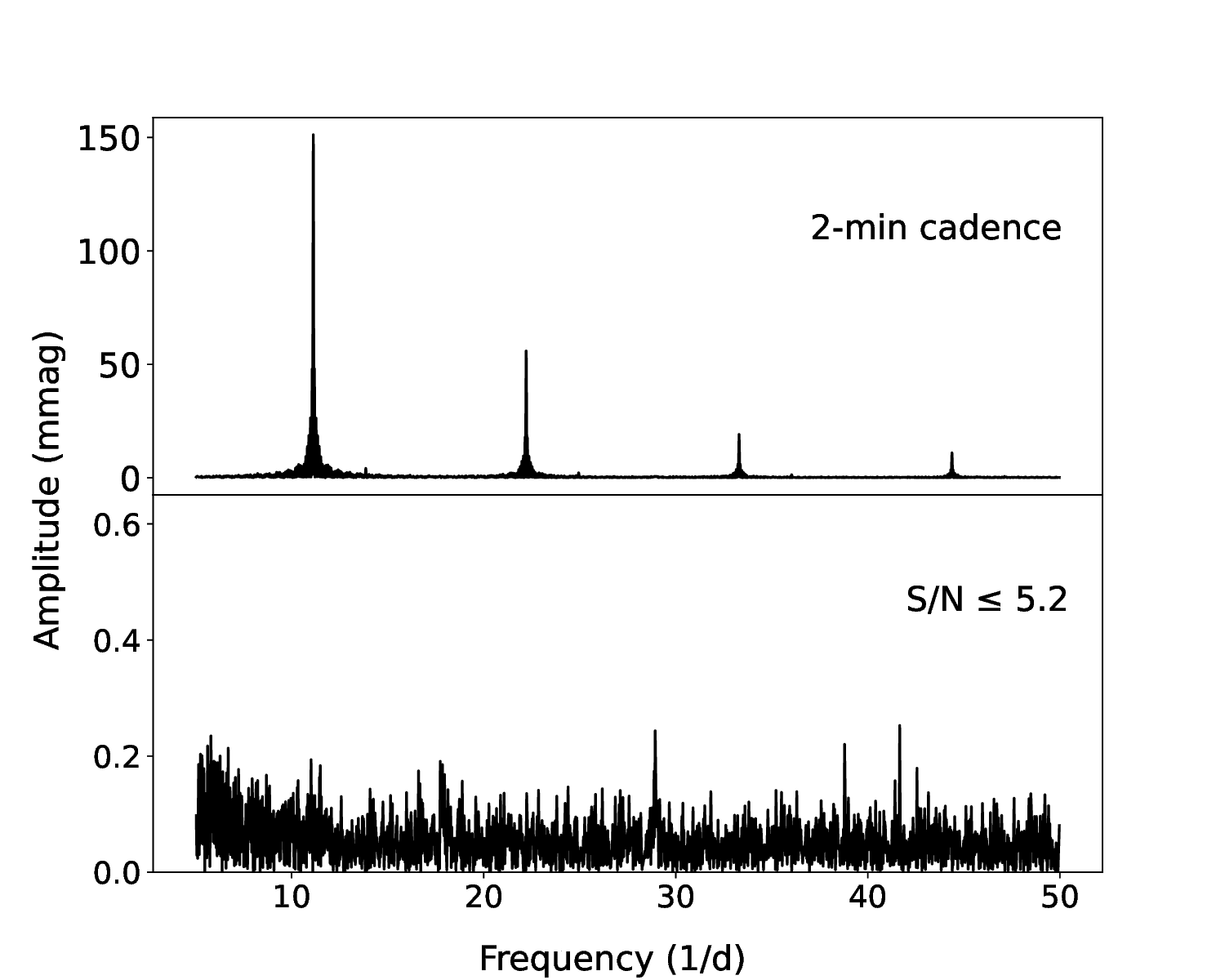}\hfill
    \caption{The short-cadence light curve of TIC 148357344 during 1.5 days (top panel), amplitude spectrum (middle panel), and amplitude spectrum of residual with S/N less than 5.2 (bottom panel).}\label{Figure 5}
\end{figure}

Figure \ref{Figure 5} displays the light curve (top panel), amplitude spectrum (middle panel), and amplitude spectrum of the residual (bottom panel) for TIC 148357344 (TYC 5968-1693-1). TIC 148357344 ($R.A. = +07^h05^m28^s.9$, $DEC = -17^{\circ}05^{\prime}16.75^{\prime\prime}$) was first identified as a variable star by \cite{Heinze2018}, then it was introduced as \dsct\ by \citet{barak2022}.
TIC 148357344 was observed only in sector 7 (short-cadence) with a duration of 27 days and consisting of 16056 data points.
We identified 12 frequencies for this star (Table \ref{tab: Table 4}), which includes three radial frequencies $\rm F0 =8.43$ $\rm day^{-1}$ ($\rm P0$ = $0.119$ days), $\rm F1 = 11.09951$ $\rm day^{-1}$ and $\rm F2 = 13.835$ $\rm day^{-1}$, as fundamental, first, and second overtones,  respectively. The frequency ratios for the first ($\rm F0/F1$) and the second ($\rm F0/F2$) overtones were obtained as 0.76 and 0.61, respectively, determining the validity of our analysis along with the {$Q$} values as 0.028, 0.023, and 0.018 for fundamental, first, and second overtones, respectively. 
Other frequencies may be known as harmonics ($f_{2}$, $f_{3}$,..., $f_{5}$) and combination frequencies ($f_{7}$, $f_{8}$, $f_{10}$, $f_{11}$, $f_{12}$).
It is noticeable that the first overtone's amplitude is more significant than that of the fundamental mode, probably owing to amplitude modulation mechanisms that are well known in \dsct\ stars \citep{Bowman2016, Lv_2021, Sun_2021}. Consequently, further studies are necessary to illuminate the driving mechanisms, including the mode-selecting processes in \dsct s.
The light curve's peak-to-peak value of about $\sim0.4$ mag and the range of frequencies propose that TIC 148357344 may be a triple-mode HADS. 

\begin{table}

	\centering
          \caption{A complete list of the 12 identified frequencies for TIC 148357344 (denoted by $f_{i}$).}\label{tab: Table 4}\
	\begin{tabular}{|c c c c c c c c|}  
		\hline
	$f_{i}$ & Frequency & Amplitude  & phase & S/N & ID & period ratio & {$Q$} value\\
		\hline
   & ($\rm day^{-1}$) &  (mmag) &  (radians/2$\pi$)&  &  &  & \\
		\hline
	1&11.09951±0.00001&151.003±0.08&0.96259 ±0.00009&1948.012&F1&0.76 &0.023±0.003\\
	2&22.19904±0.00003&55.44±0.08&0.3230±0.0002&942.301&2F1& -&-\\
	3&33.29853±0.00009&19.09±0.08&0.6087±0.0002&334.345&3F1&- &-\\
        4&44.3980±0.0001&10.92±0.08&0.8494±0.0007&198.396&4F1& -&-\\
        5&55.4977±0.0006&6.01±0.08&0.483±0.001&134.900&5F1&- &-\\
        6&13.835±0.001&2.93±0.08&0.592±0.004&55.0831&F2& 0.61 &0.018±0.002\\
        7&24.935±0.001&1.85±0.08&0.426±0.007&34.443&F1+F2& -&-\\
        8&47.132±0.003&0.56±0.08&0.67±0.01&11.070&3F1+F2& -&-\\
        9&8.43±0.003&0.51±0.08&0.98±0.02&12.999&F0& -&0.028±0.004\\
        10&58.233±0.003&0.41±0.08&0.22±0.02&11.216&4F1+F2&-&-\\
        11&30.559±0.004&0.39±0.08&0.610±0.003&9.018&4F1-F2& -&-\\
        12&19.463±0.004&0.38±0.08&0.717±0.003&6.32&3F1-F2& -&-\\
		\hline 
  \multicolumn{2}{l}{ Note. The frequency resolution is about $f_{res}=0.061 \rm day^{-1}$.}	
	\end{tabular}
  
\end{table}

\subsection{TIC 160120432}
\begin{figure}
\centering
    \includegraphics[width=.7\textwidth]{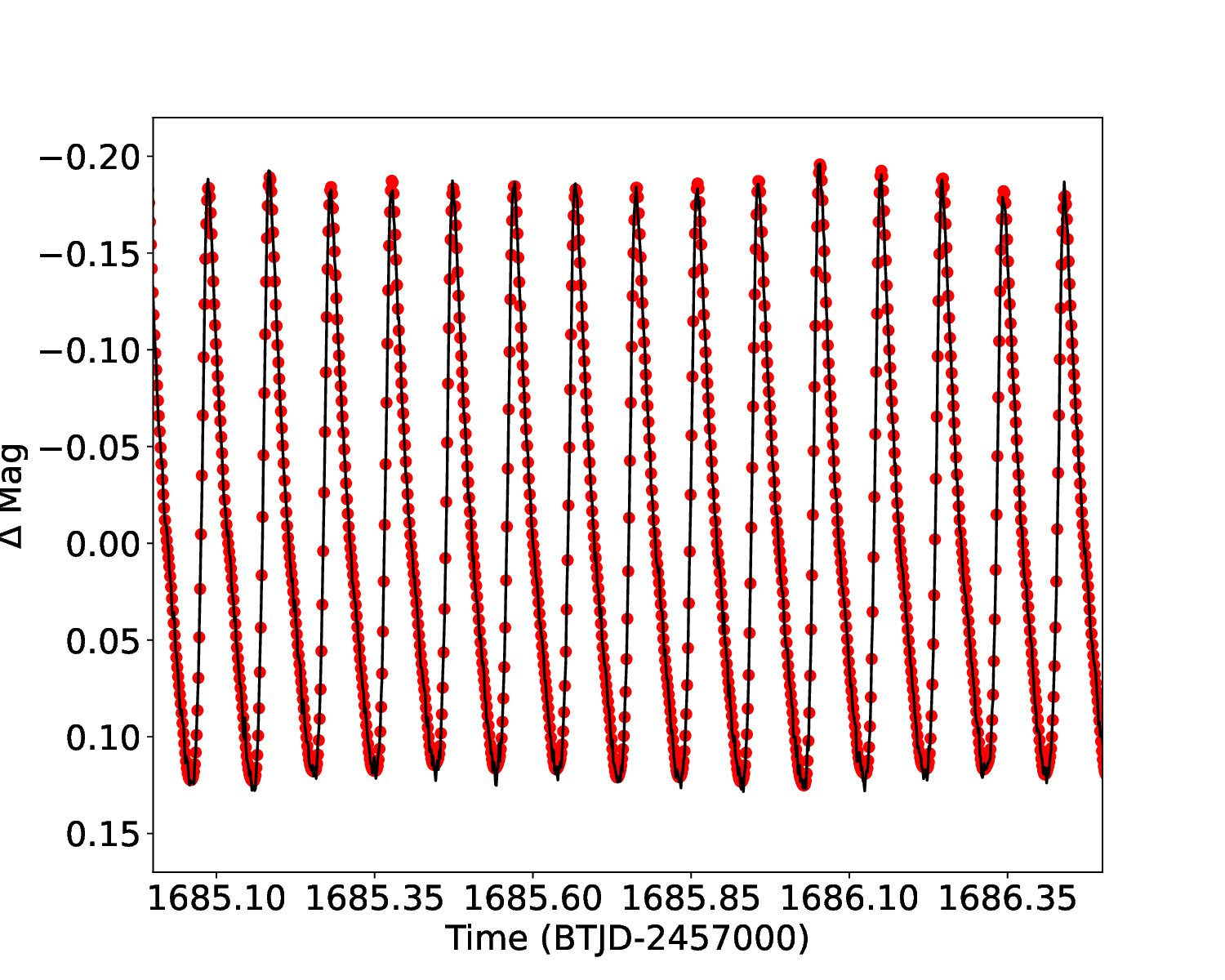}\hfill
    \\[\smallskipamount]
    \includegraphics[width=.7\textwidth]{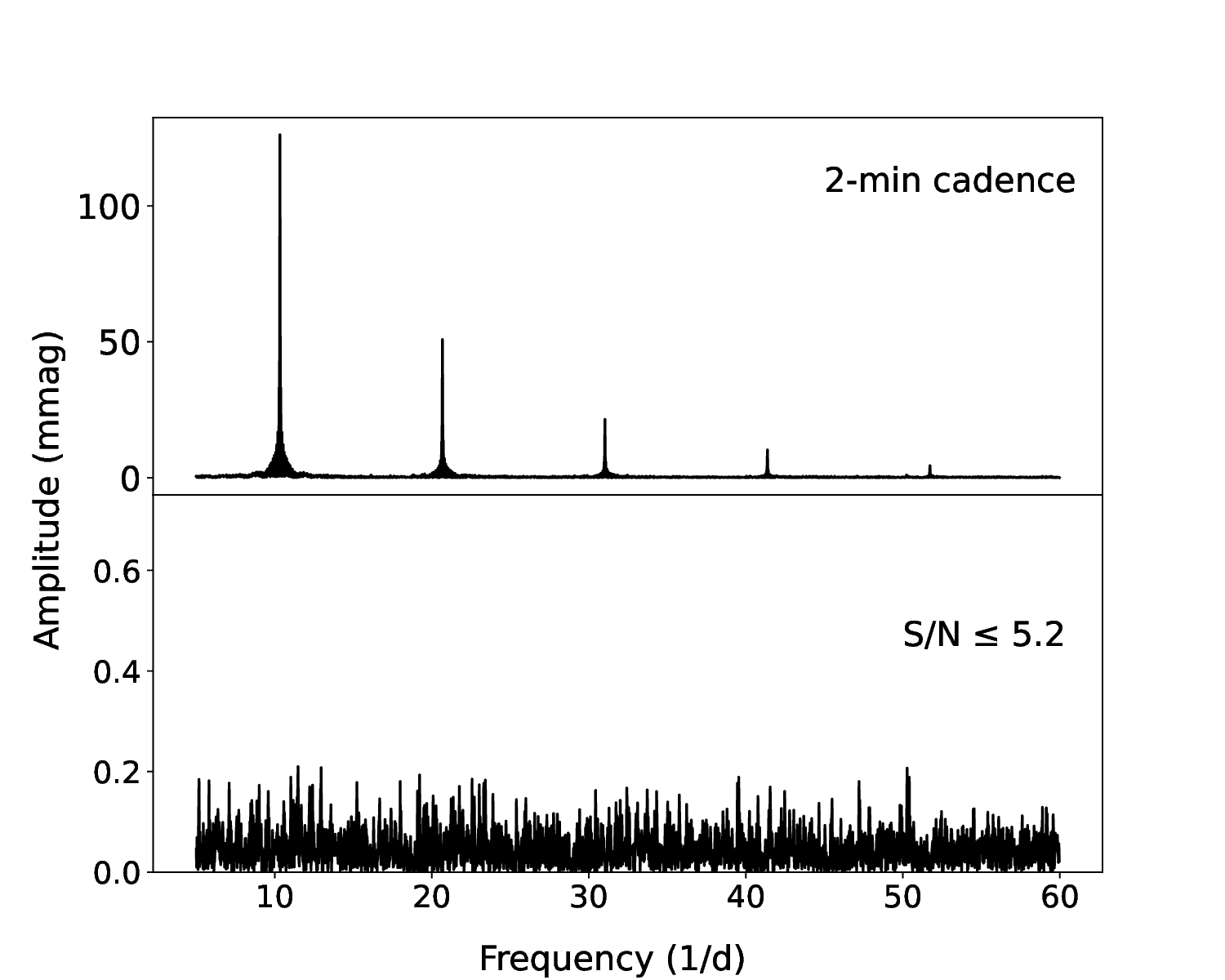}\hfill
   \caption{The short-cadence light curve (sector 14) of TIC 160120432 during 1.5 days (top panel), amplitude spectrum (middle panel), and amplitude spectrum of residual with S/N less than 5.2 (bottom panel).}\label{Figure 6}
\end{figure}

Figure \ref{Figure 6} shows the light curve (top panel), amplitude spectrum (middle panel), and amplitude spectrum of the residual (bottom panel) for TIC 160120432 (TYC 4638-455-1). Using NSVS and ASAS-3 databases, \citet{Otero2007} suggested TIC 160120432 ($R.A. = +15^h35^m30^s.40$, $DEC = +85^{\circ}37^{\prime}38.94^{\prime\prime}$) as a HADS star.
We applied frequency analysis for combined sectors (19 and 20), individual sectors (14, 40, and 47), and combined sectors (52 and 53). We pick up the significant frequencies (S/N$> 5.2$ and frequency resolution greater than 0.028 $\rm day^{-1}$ and 0.061 $\rm day^{-1}$ for combined and individual sectors, respectively).
In Table \ref{tab: Table 5}, we found 25 frequencies for TIC 160120432, which include two radial frequencies $\rm F0= 10.350856$ $\rm day^{-1}$ ($\rm P0$ = $0.0966$ days) and $\rm F1 = 13.612$ $\rm day^{-1}$, with {$Q$} values as 0.031 and 0.023, respectively.
As shown in Table\ref{tab: Table 5}, we specified the harmonic frequencies ($f_{2}$, $f_{3}$,..., $f_{5}$), combination frequencies ($f_{7}$, $f_{11}$, $f_{12}$, $f_{14}$), and nonradial modes ($f_{6}$, $f_{8}$, $f_{9}$, $f_{13}$, $f_{15}$..., $f_{25}$).
The light curve's peak-to-peak value is about $\sim0.33$ mag. The frequencies of TIC 160120432 are ranging from 9 to 51 $\rm day^{-1}$. The frequency ratio ($\rm F0/F1$) is 0.76, and the {$Q$} values were calculated as 0.031 and 0.024 for the fundamental and first overtone, respectively. These four findings strongly verified the HADS property for TIC 160120432 as a new double-mode HADS star.

\begin{table}
    
	\centering
         \caption{A complete list of the 25 identified frequencies for TIC 160120432 (denoted by $f_{i}$).}\label{tab: Table 5}\
         
          \begin{tabular}{|c c c c c c c c|}  
		\hline
	$f_{i}$ & Frequency  & Amplitude  & phase & S/N & ID & period ratio & {$Q$} value\\
		\hline
   &  ($\rm day^{-1}$) & (mmag) &  (radians/2$\pi$)&  &  &  & \\
   \hline
	1 & 10.350856±0.000003 & 126.46±0.05 & 0.213±0.00006 & 899.46 & F0 & - & 0.031±0.00323\\
        2 & 20.701511±0.000007 & 50.991±0.05 & 0.400±0.0001 & 424.95 & 2F0 & - & -\\
	3 & 31.05230±0.00001 & 21.50±0.05 & 0.394±0.0003& 245.05 & 3F0 &  -& -\\
        4 & 41.40302±0.00003 & 10.378±0.05 & 0.710±0.0008 & 168.428 & 4F0 & - & -\\
        5 & 51.75381±0.00008 & 4.86±0.05 & 0.051±0.001 & 102.31 & 5F0 & - & -\\
        6 & 9.6819±0.0006 & 1.167±0.05 & 0.503±0.007 & 10.71 &nonradial&  -& -\\
        7 & 19.5715±0.0007 & 0.956±0.05 & 0.833±0.008 & 11.01 &6F1-6F0& - & -\\
        8 & 16.1607±0.0009 & 1.069±0.05 & 0.641±0.007 & 15.85 & nonradial& - &- \\
        9 & 18.8482±0.0001 & 0.912±0.05 & 0.754±0.002 & 9.24 & nonradial & - &-\\
        10 & 13.612±0.001 & 0.219±0.05 & 0.062±0.03 & 8.880 & F1 & 0.76 & 0.023±0.003\\
        11 & 29.9194±0.0008 & 0.625±0.05 & 0.249±0.01 & 9.96 &6F1-5F0&-  &- \\
        12 & 9.219±0.001 & 0.546±0.05 & 0.626±0.01 & 6.89 & 6F1-7F0 & - &- \\
        13 & 26.510±0.001 & 0.524±0.05 & 0.745±0.01 & 9.74& nonradial & - &- \\
        14 & 40.269±0.001 & 0.487±0.05& 0.836±0.01 & 7.45 &6F1-4F0& - & -\\
        15 & 19.6427±0.0009 & 0.555±0.05 & 0.416±0.01 & 8.24 & nonradial & - &- \\
        16 & 40.0626±0.0009 & 0.511±0.05 & 0.167±0.01 & 9.80 & nonradial& - & -\\
        17 & 19.4962±0.0005 & 0.539±0.05 & 0.876±0.01 & 7.772 &nonradial  &-  &- \\
        18 & 30.380±0.001 & 0.458±0.05 & 0.680±0.01 & 7.127 & nonradial & - & - \\
	20 & 18.8807±0.0002 & 0.531±0.05 & 0.280±0.02 & 8.329 & nonradial & - & -\\
  	21 & 29.8476±0.0006 & 0.57±0.05 & 0.002±0.02 & 9.66 & nonradial & - & -\\
	22 & 22.0570±0.0008 & 0.48±0.05 & 0.22±0.02 & 9.64 & nonradial& - & -\\
	23 & 40.336±0.001 & 0.35±0.05 & 0.74±0.02 & 7.47 & nonradial & - & -\\
	24 & 15.520±0.001 & 0.34±0.05 & 0.25±0.02 & 7.51 & nonradial & - & -\\
	25 & 9.737±0.001 & 0.36±0.05 & 0.31±0.02 & 6.56 & nonradial & - & -\\

		\hline
	\end{tabular}

\end{table}

\subsection{TIC 278119167}
\begin{figure}
\centering
     \includegraphics[width=.7\textwidth]{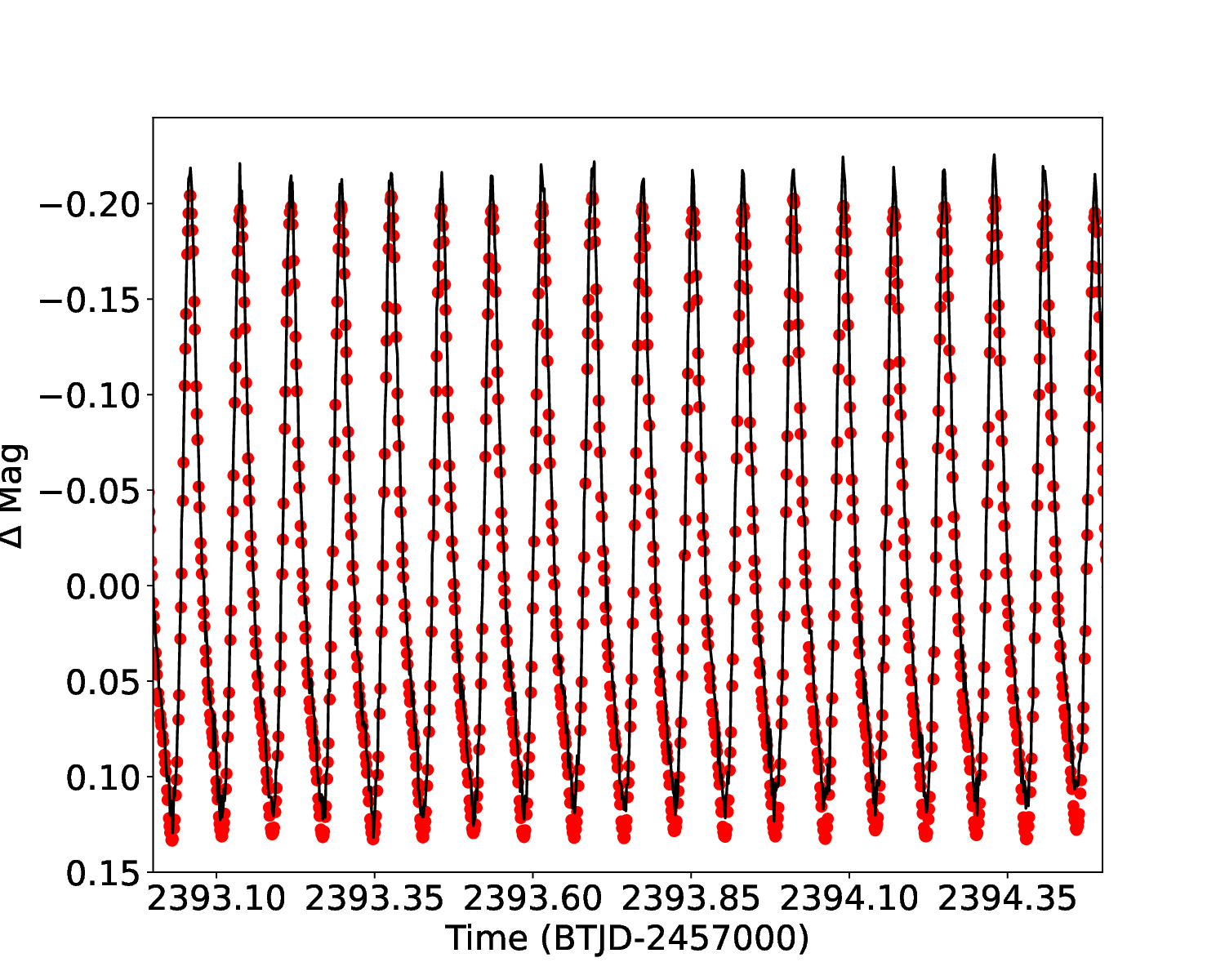}\hfill
    \\[\smallskipamount]
    \includegraphics[width=.7\textwidth]{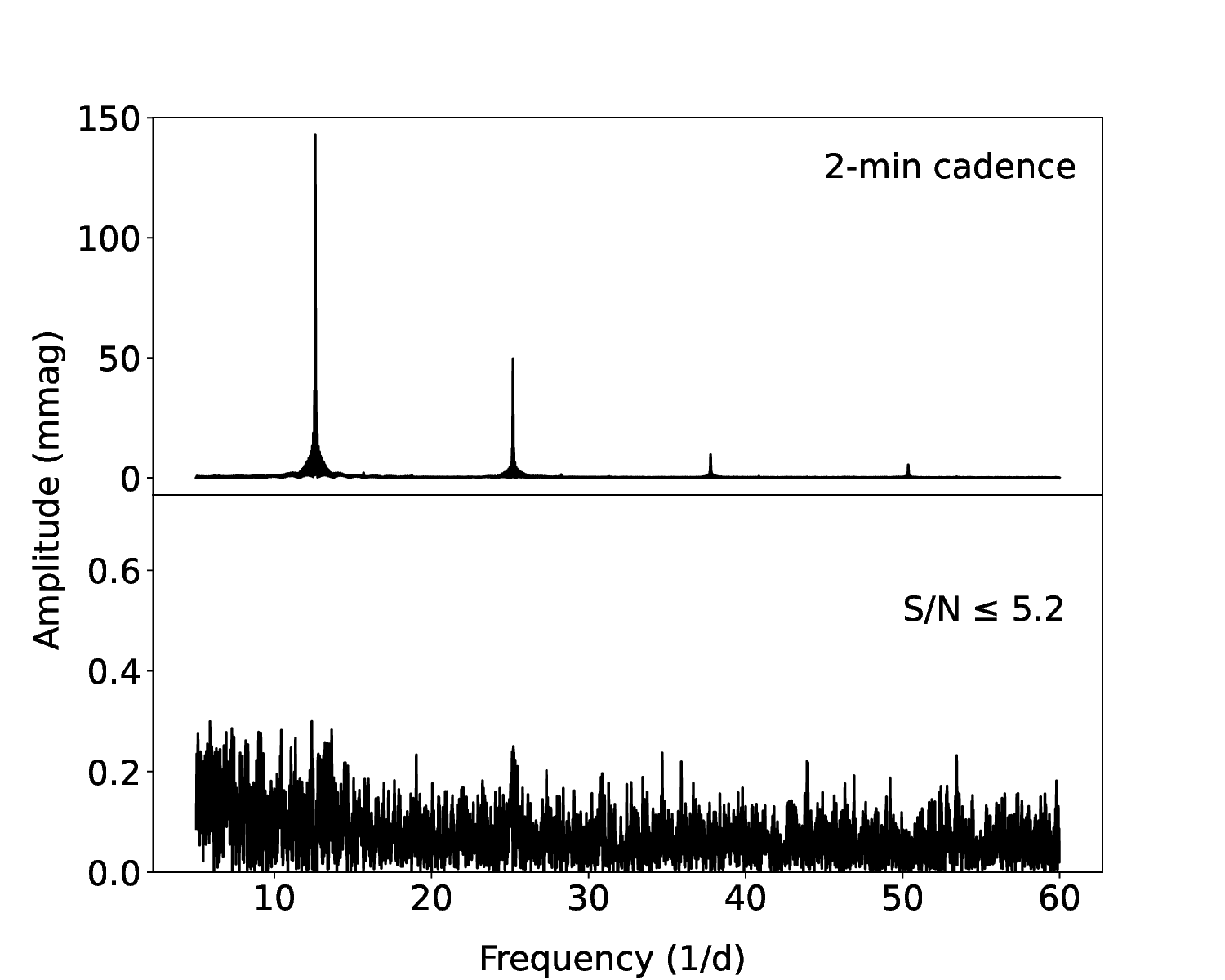}\hfill 
    \caption{The short-cadence light curve (sector 40) of TIC 278119167 during 1.5 days (top panel), amplitude spectrum (middle panel), and amplitude spectrum of residual with S/N less than 5.2 (bottom panel).}\label{Figure 7}
\end{figure}

Figure \ref{Figure 7} represents the light curve (top panel), amplitude spectrum (middle panel), and amplitude spectrum of the residual (bottom panel) for TIC 278119167 (TYC 1578-1754-1). \cite{Heinze2018} categorized TIC 278119167 ($R.A. = 18^h38^m50^s.36$, $DEC = +19^{\circ} 07^{\prime} 44.61^{\prime\prime}$) as a pulsating variable component of a contact/overcontact binary system using a probabilistic machine-learning algorithm. Our investigation indicates that TIC 278119167 is triple-mode HADS.
We analyzed the light curves for two individual sectors, 40 and 53, with a duration of 28.1 and 24.98 days, respectively. Interestingly, the set of significant frequencies for the light curve of both sections is approximately the same. We identified nine frequencies constructing the star's light curve (Table \ref{tab: Table 6}), which includes three radial frequencies $\rm F0 = 9.579$ $\rm day^{-1}$ ($\rm P0$ = $0.105$ days), $\rm F1 = 12.58960$ $\rm day^{-1}$ and $\rm F2 = 15.671$ $\rm day^{-1}$, with {$Q$} values as 0.031, 0.023, and 0.018 respectively. 
Other frequencies may be represented as harmonics ($f_{2}$, $f_{3}$, $f_{4}$, $f_{9}$) and combination frequencies ($f_{6}$, $f_{7}$).
Again, in this star, the amplitude modulation phenomenon is probable due to the first overtone's amplitude being more significant than the fundamental frequency's amplitude \citep{Bowman2016, Lv_2021, Sun_2021}. Consequently, further studies are necessary to illuminate the driving mechanisms and mode-selecting processes, which are still not fully understood in \dsct\ stars.
The rectified light curve shows a peak-to-peak amplitude of about $\sim0.43$ mag, which may be suggested as a new triple-mode HADS.

\begin{table}
    	\centering
         \caption{A complete list of the Nine identified frequencies for TIC 278119167 (denoted by $f_{i}$). }\label{tab: Table 6}\
         
          \begin{tabular}{|c c c c c c c c|} 
		\hline
	$f_{i}$ & Frequency  & Amplitude  & phase & S/N & ID & period ratio & {$Q$} value\\
		\hline
		 &  ($\rm day^{-1}$) &  (mmag) &  (radians/2$\pi$)&  &  && \\
		\hline1&12.58960±0.00002&176.2±0.1&0.5448±0.0002&797.74&F1&0.76&0.023±0.002\\
	2&25.17914±0.00007&49.6±0.1&0.1273±0.0006&565.91&2F1&-&-\\
	3&37.7685±0.0003&9.7±0.1&0.051±0.003&142.05&3F1&-&-\\
        4&50.3583±0.0006&5.3±0.1&0.745±0.005&95.015&4F1&-&-\\
        5&15.671±0.002&1.5±0.1&0.46±0.01&15.85&F2&0.61&0.018±0.002\\
        6&18.746±0.002&1.4±0.1&0.19±0.02&18.098&2F2-F1&-&-\\
        7&28.260±0.003&1.0±0.1&0.48±0.02&16.202&F1+F2&-&-\\
        8&9.579±0.005&0.2±0.1&0.47±0.04&5.31&F0&-&0.031±0.003\\
        9&31.323±0.005&0.7±0.1&0.64±0.08&11.72&2F2&-&-\\
		\hline
  \multicolumn{2}{l}{ Note. The frequency resolution is about $f_{res}=0.053 \rm day^{-1}$.}
	\end{tabular}
\end{table}
\subsection{TIC 710783}
\begin{figure}
\centering
 \includegraphics[width=.7\textwidth]{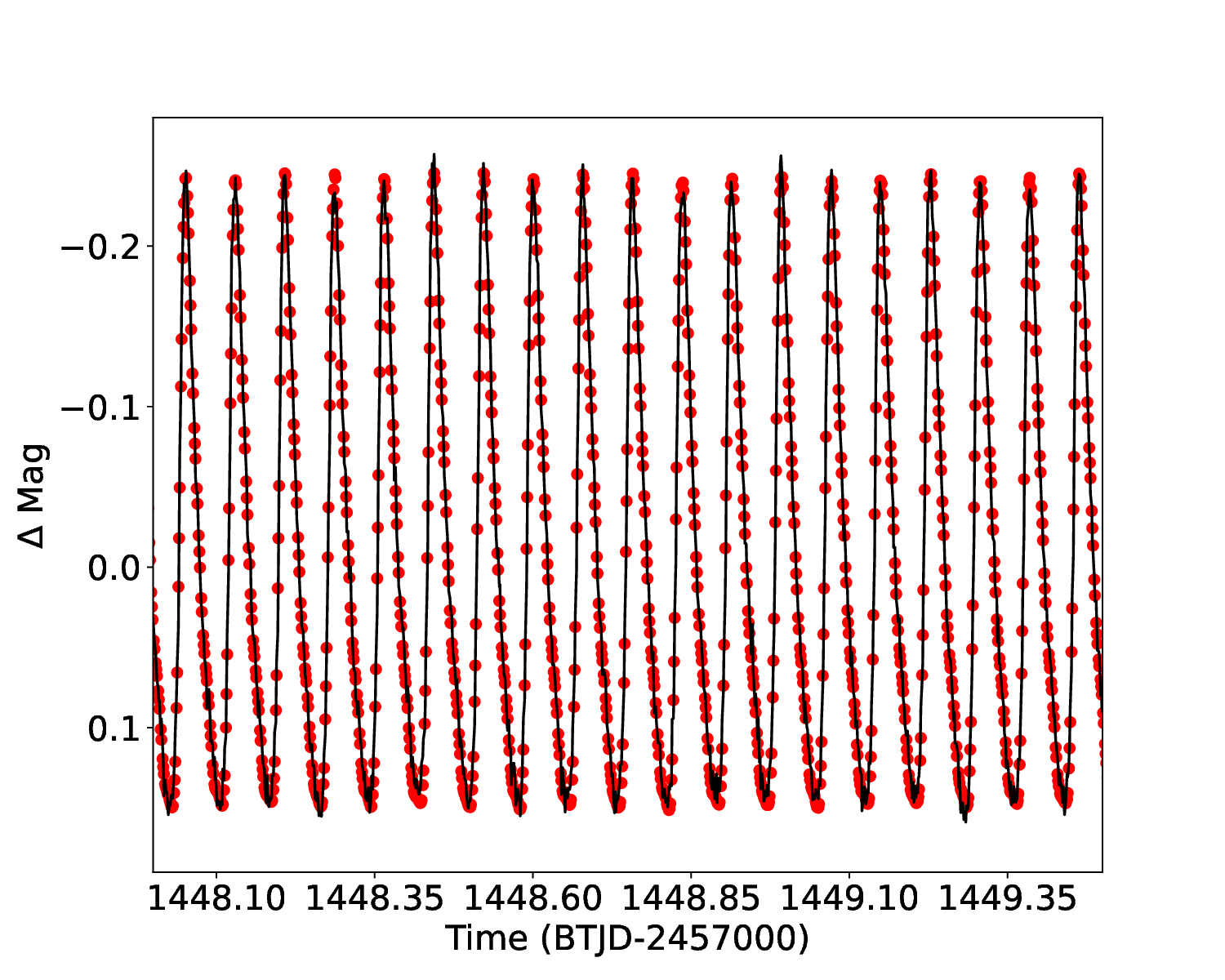}\hfill
    \\[\smallskipamount]
   \includegraphics[width=.7\textwidth]{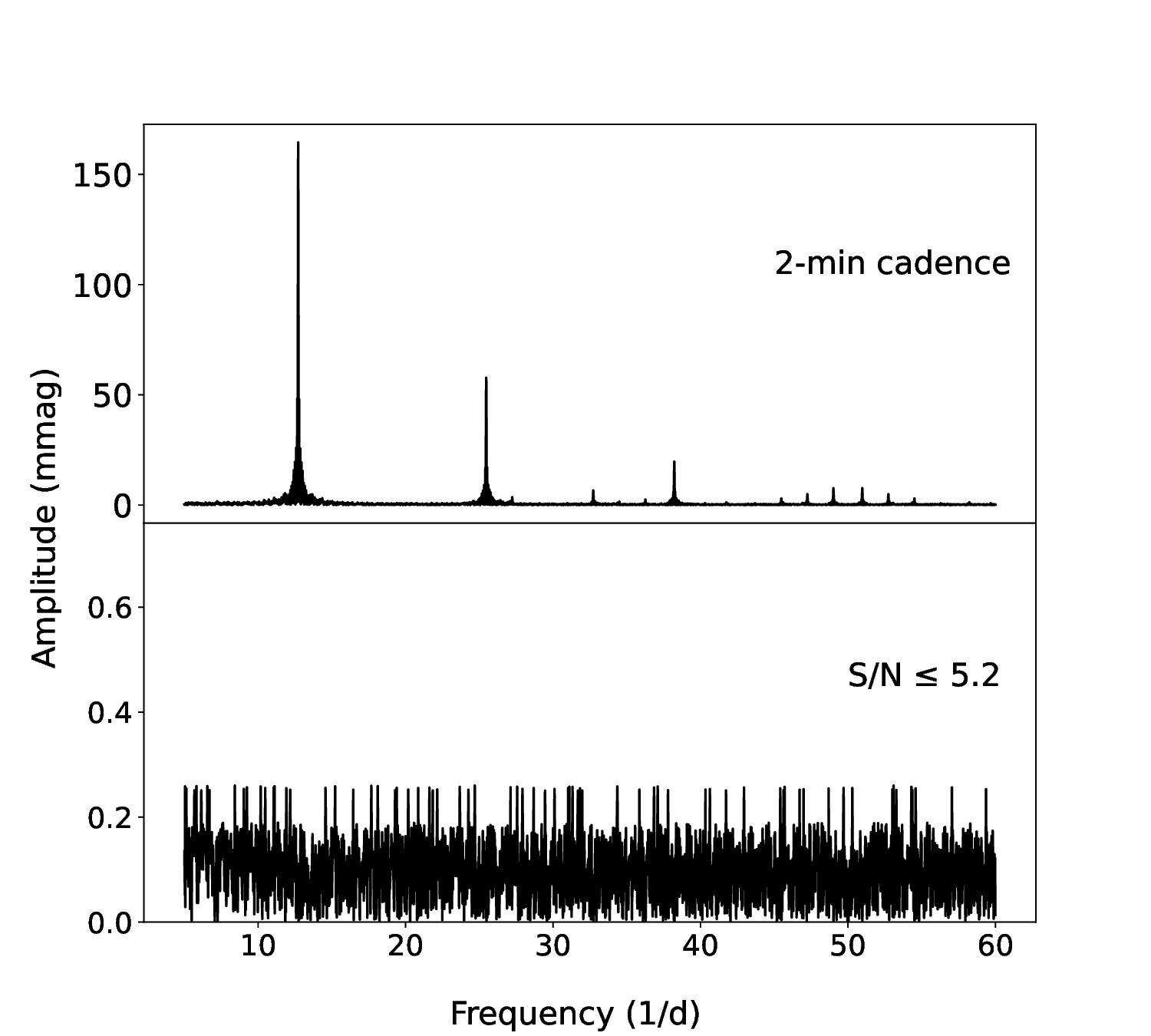}\hfill
    \caption{The short-cadence light curve of TIC 710783 during 1.5 days (top panel), amplitude spectrum (middle panel), and amplitude spectrum of residual with S/N less than 5.2 (bottom panel).}\label{Figure 8}
\end{figure}

Figure \ref{Figure 8} represents the light curve (top panel), amplitude spectrum (middle panel), and amplitude spectrum of the residual (bottom panel) for TIC 710783 (ATO J074.1485-27.6801). 
The TIC 710783 ($R.A. = 04^h56^m35^s.63$, $DEC =-27^{\circ}40^{\prime}48.56^{\prime\prime}$) was first recognized as a variable star by \cite{Heinze2018}, and then its physical parameters (temperature, gravity, and frequency scaling relation) were measured by \cite{2020_barcelo} as a \dsct\ star. We suggest that TIC 710783 shows signs of a single-mode HADS.
The \tess\ space telescope recorded TIC 710783 during sector 5 for 25.43 days and consisting 17,023 data points.
We identified 11 frequencies constructing the star's light curve (Table \ref{tab: Table 7}), which includes a radial frequency $\rm F0 = 12.74451$ $\rm day^{-1}$ ($\rm P0$ = $0.078$ days), with {$Q$} value as 0.039. We observed the harmonics frequencies ($f_{2}$, $f_{3}$, $f_{4}$) and nonradial modes ($f_{5}$,$f_{6}$,...,$f_{11}$).
The light curve's peak-to-peak value of about $\sim0.47$ mag and the range of frequencies propose that TIC 710783 may be a new monomode HADS. 

\begin{table}
    	\centering
         \caption{A complete list of the 11 identified frequencies for TIC 710783 (denoted by $f_{i}$).}\label{tab: Table 7}\
      
          \begin{tabular}{|c c c c c c c c|}  
		\hline
	$f_{i}$ & Frequency  & Amplitude  & phase & S/N & ID & period ratio & {$Q$} value\\
		\hline
   &  ($\rm day^{-1}$) &  (mmag) &  (radians/2$\pi$)& &  & & \\
		\hline
		1 & 12.74451±0.00005 & 164.6±0.4 & 0.4635±0.0003 & 738.43& F0 & - & 0.039±0.003 \\
		2 & 25.4890±0.0001 & 57.8±0.4 & 0.828±0.001 & 350.83& 2F0 & - &- \\
		3 & 38.2334±0.0004  & 19.7±0.4 & 0.29±0.03 & 126.32& 3F0 &- &- \\
        4 & 50.977±0.001 & 7.8±0.4 & 0.922±0.008 & 68.48 & 4F0 & - & -\\
        5 & 7.255±0.003 & 2.8±0.4& 0.89±0.02 &10.83 & nonradial &- & -\\
        6 & 36.277±0.003 & 2.7±0.4 & 0.18±0.02 & 22.87 & nonradial & - &-\\
        7 & 5.489±0.003 & 2.4±0.4 & 0.73±0.02 & 7.88 & nonradial& - &-\\
        8 & 32.745±0.003 & 2.3±0.4 & 0.37±0.02 & 18.55 &nonradial & -&- \\
        9 & 54.508±0.006 & 1.4±0.4 & 0.63±0.04 & 11.05 & nonradial &-  &- \\
        10 & 18.231±0.008 & 1.0±0.4 & 0.74±0.06 & 8.35 & nonradial &-  & -\\
        11 & 11.838±0.009 & 0.9±0.4 & 0.76±0.07 & 5.44 & nonradial & - & -\\
		\hline
  \multicolumn{2}{l}{ Note. The frequency resolution is about $f_{res}=0.058 \rm day^{-1}$.}
	\end{tabular}
	

\end{table}
\subsection{TIC 187386415}
\begin{figure}
    \centering
    \includegraphics[width=.7\textwidth]{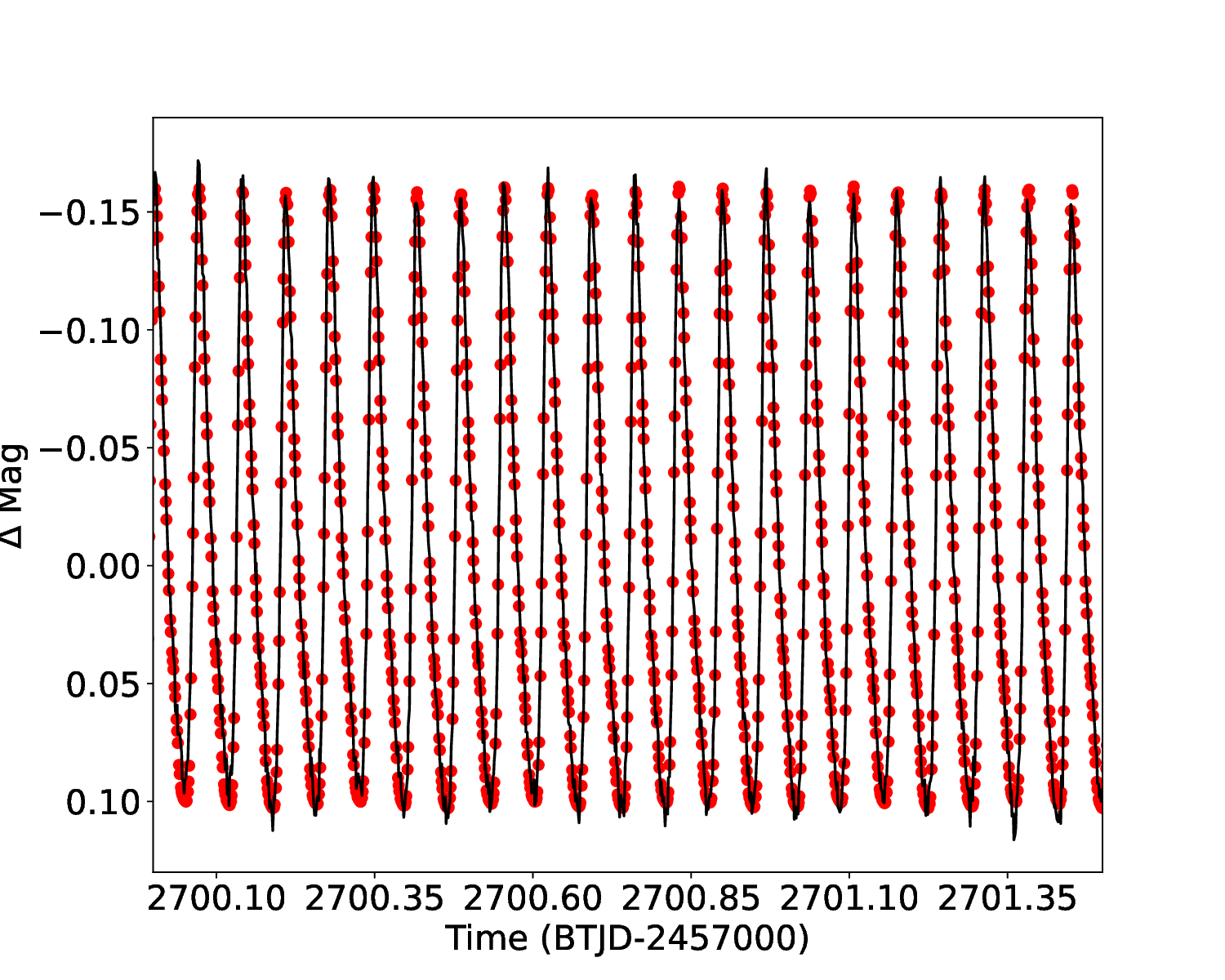}\hfill
    \\[\smallskipamount]
    \includegraphics[width=12cm,height=10cm]{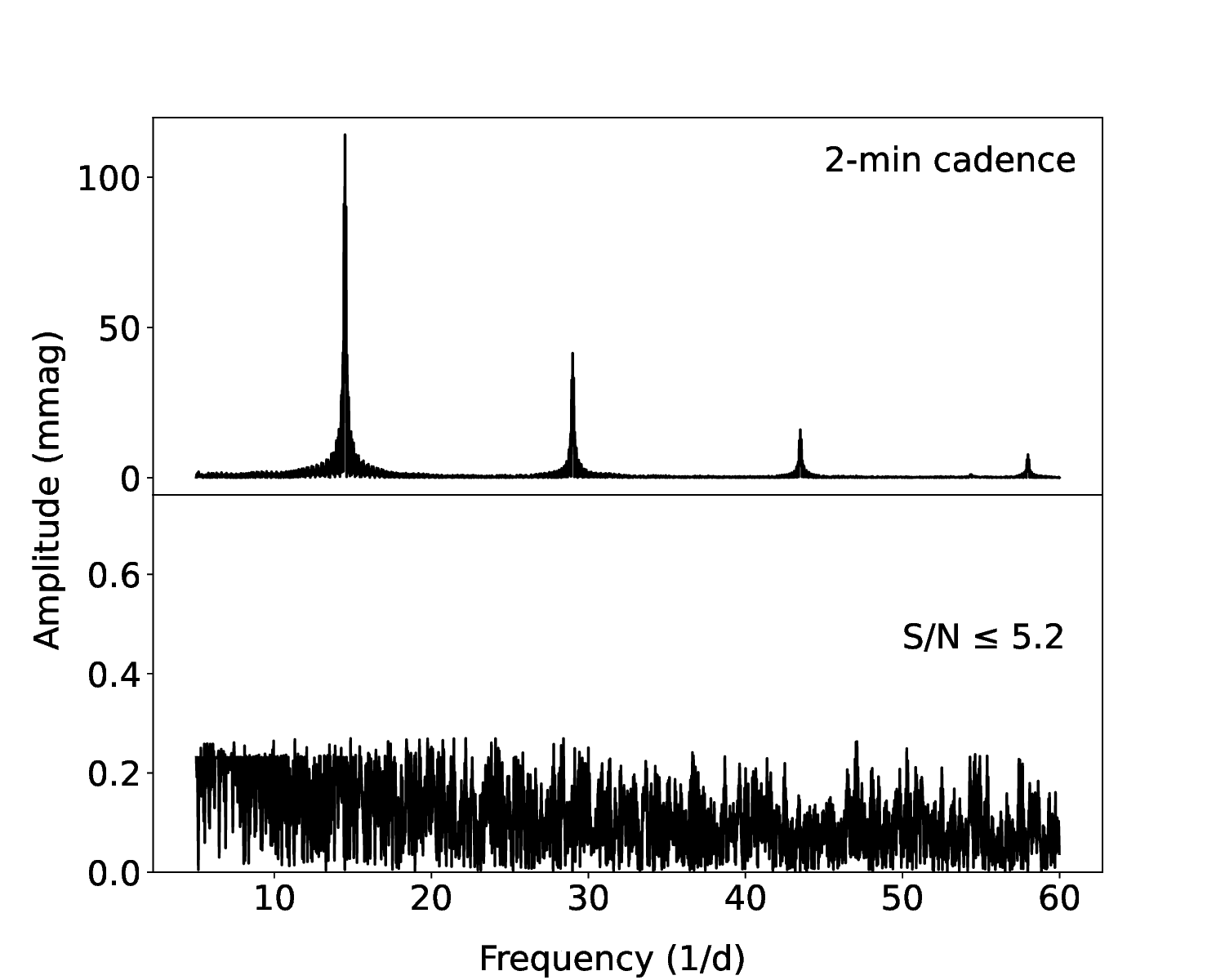}\hfill
    \caption{The short-cadence light curve of TIC 187386415 during 1.5 days (top panel), amplitude spectrum (middle panel), and amplitude spectrum of residual with S/N less than 5.2 (bottom panel).}\label{Figure 9}
\end{figure}

Figure \ref{Figure 9} represents the light curve (top panel), amplitude spectrum (middle panel), and amplitude spectrum of the residual (bottom panel) for TIC 187386415 (ATO J236.5561-00.4351). TIC 187386415 ($R.A. = 15^h46^m13^s.48$, $DEC =-00^{\circ}26^{\prime}06.34^{\prime\prime}$) was classified by \cite{Hey2021} as a variable star. This study shows that TIC 187386415 is a single-mode HADS. 
We analyzed this star for sector 51 with a duration of 19.13 days and consisting of 6775 data points.
We identified four frequencies constructing the star's light curve (Table \ref{tab: Table 8}), which includes a radial frequency $\rm F0 = 14.49465$ $\rm day^{-1}$ ($\rm P0$ = $0.069$ days), with {$Q$} value as 0.033. Other frequencies may be introduced as harmonic frequencies ($f_{2}$, $f_{3}$, $f_{4}$).
The peak-to-peak amplitude ($\sim0.31$ mag) and the frequencies' range propose that TIC 187386415 may be a new monomode HADS star.

\begin{table}
	\centering
        \caption{A complete list of the Four identified frequencies for TIC 187386415 (denoted by $f_{i}$).}\label{tab: Table 8}\
         
	\begin{tabular}{|c c c c c c c c|}  
		\hline
	$f_{i}$ & Frequency & Amplitude & phase & S/N & ID & period ratio & {$Q$} value\\
		
  \hline
	 &  ($\rm day^{-1}$) &  (mmag) & (radians/2$\pi$)&  &  &  & \\
		\hline
		1&14.49465±0.00003&113.8±0.1&0.5514±0.0001&473.8&F0&-&0.033±0.0045\\
		2&28.98936±0.00009&41.0±0.1&0.8682±0.0005&333.91&2F0&-&-\\
		3&43.4840±0.0002&15.6±0.1&0.223±0.001&196.24&3F0&-&-\\
        4&57.9786±0.0005&7.5±0.1&0.989±0.003&93.54&4F0&-&-\\
		\hline 
  \multicolumn{2}{l}{Note. The frequency resolution is about $f_{res}=0.078 \rm day^{-1}$.}
	
	\end{tabular}
 
\end{table}

\subsection{Asteroseismical parameters vs. physical quantities}\label{subsec:Asteroseismical parameters}
\begin{figure}
\centering
 \includegraphics[width=.5\textwidth]{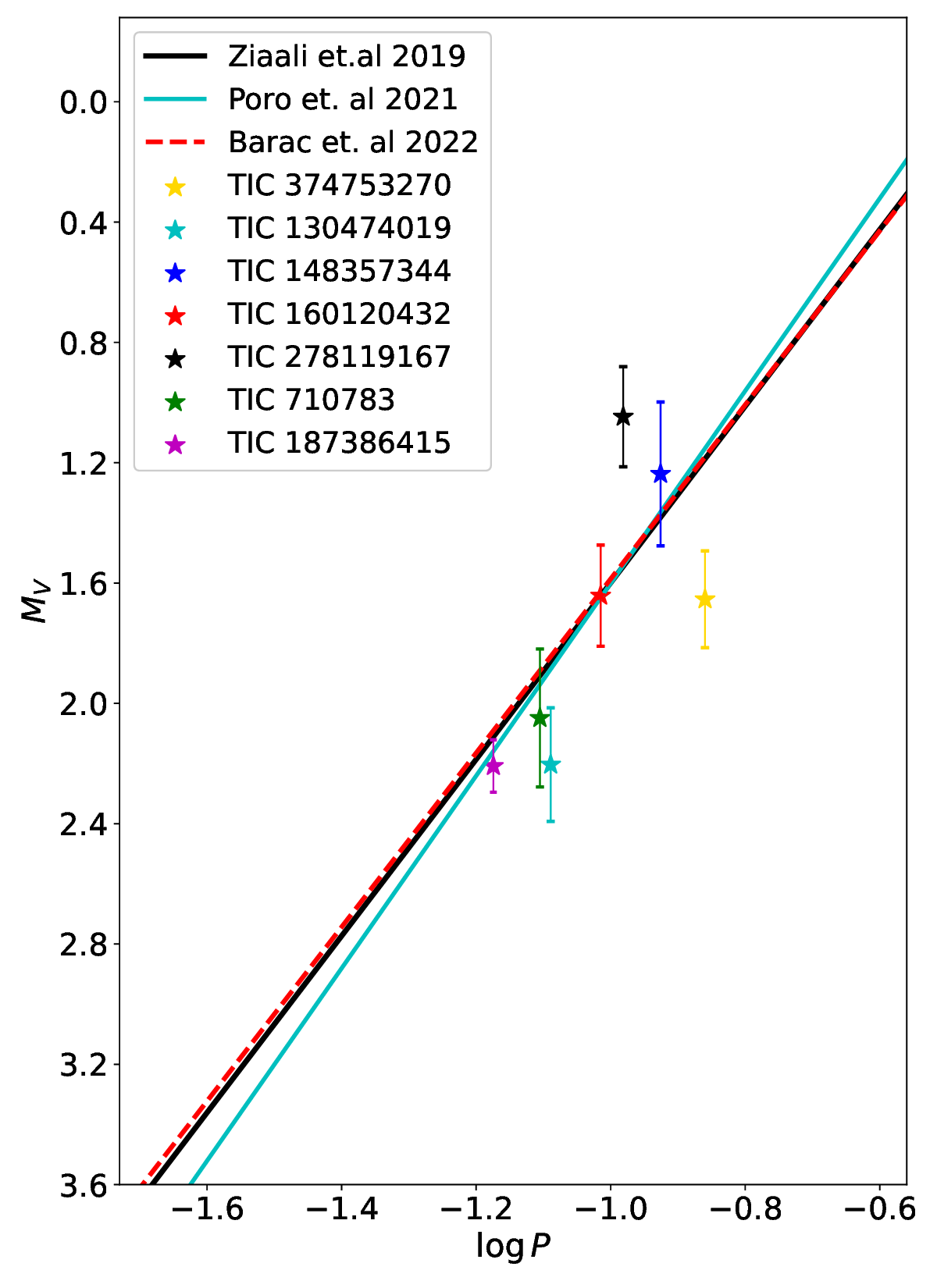}\hfill
   \includegraphics[width=.5\textwidth]{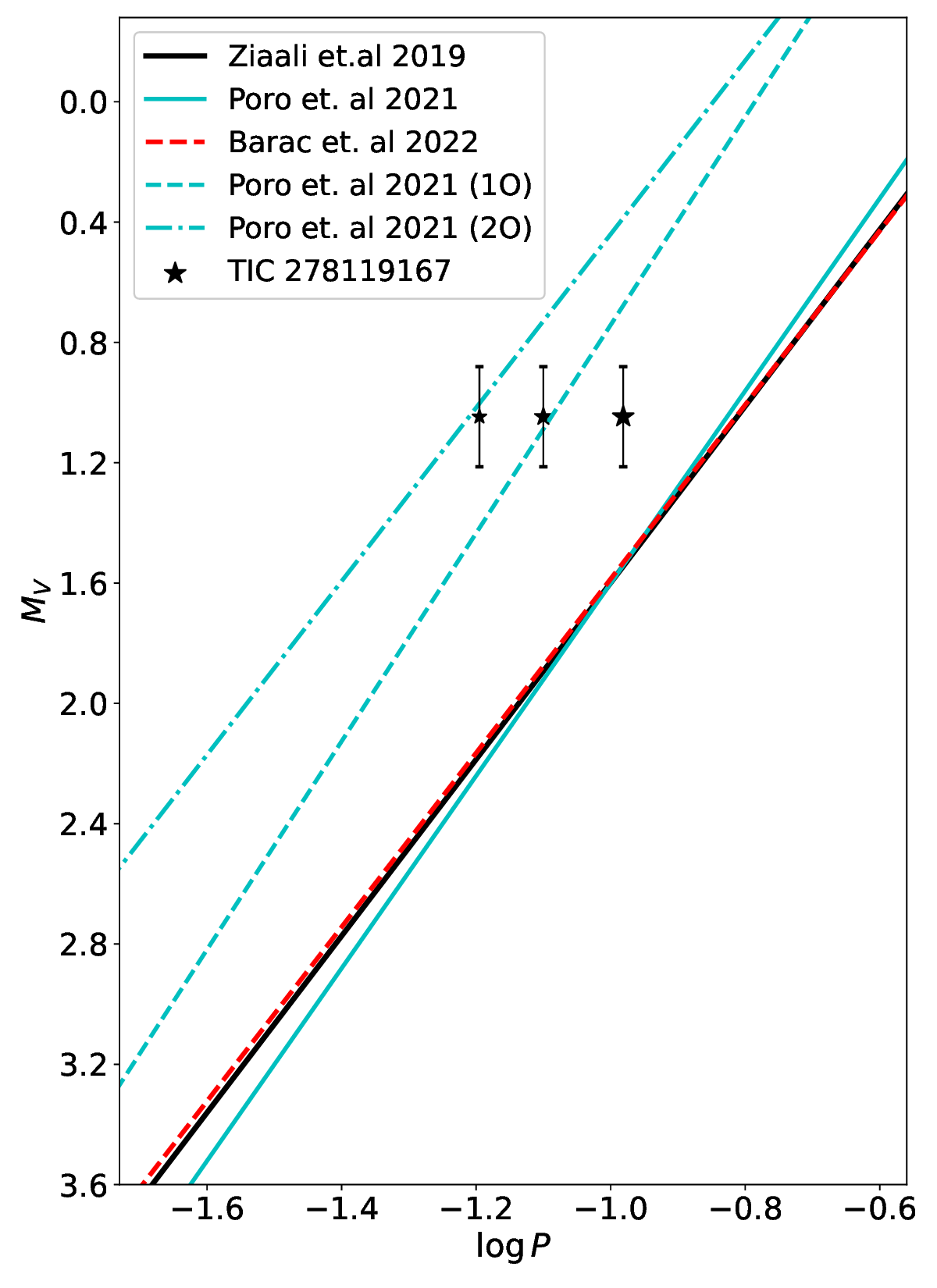}\hfill
   \\[\smallskipamount]
    \caption{(Left panel) The period-luminosity (PL) diagram of our HADS stars (colored star markers).
    The solid black line shows the relation from \cite{Ziaali_2019} (see Equation \ref{eq.5}). The diagonal dashed red line shows the relation from \citet{barak2022} (see Equation \ref{eq.6}). The cyan dashed–dotted line shows the relation from \citet{Poro_2021} (see Equation \ref{eq.7}). The error bars indicate the errors for absolute magnitudes.
    (Right panel) The PL relation of TIC 278119167 for the first (cyan dashed line) and second (cyan dotted line) overtones. While the fundamental mode (the largest star marker) slightly deviates from fundamental lines, the first (the middle-sized star marker) and second (the smallest star marker) overtones are entirely consistent with their PL relations.}\label{Figure 10}
\end{figure}

Using  Equation (\ref{eq.3}), we obtained the absolute magnitudes and their errors (applying errors of apparent magnitudes and parallaxes) for seven HADS stars.
We calculated the period ($\rm P0$) of each star using the fundamental frequency (Tables \ref{tab: Table 2}-\ref{tab: Table 8}).
Figure \ref{Figure 10} (left panel) shows the PL relation for our HADS stars (colored star markers), the PL relation of Equation (\ref{eq.5}) (solid black), Equation (\ref{eq.6}) (red dashed line), and Equation (\ref{eq.7}) (cyan dashed–dotted line). For stars TIC 374753270 and TIC 130474019, the fundamental period and absolute magnitude agreed with \cite{barak2022}. 
As shown in Figure \ref{Figure 10}, all seven cases satisfy the PL relation for \dsct\ stars. that indicates the validity of our frequency analysis.

Figure \ref{Figure 10} (right panel) examines the first and second overtone periods of TIC 278119167 by PL relations for the first (dashed line) and second (dashed–dotted line) overtones. The PL relation for the first and second overtones was obtained by \citet{Poro_2021}. While the TIC 278119167 star's fundamental period (shown as the largest star marker) slightly deviates from the fundamental PL relation, the first overtone period (the middle-sized star marker) and the second overtone period (the smallest star marker) entirely agree with the first and second overtone PL relations. This implies the validity of the identified fundamental frequency for TIC 278119167.

\begin{figure}
    \centering
    \includegraphics[width=11.cm,height=9.cm]{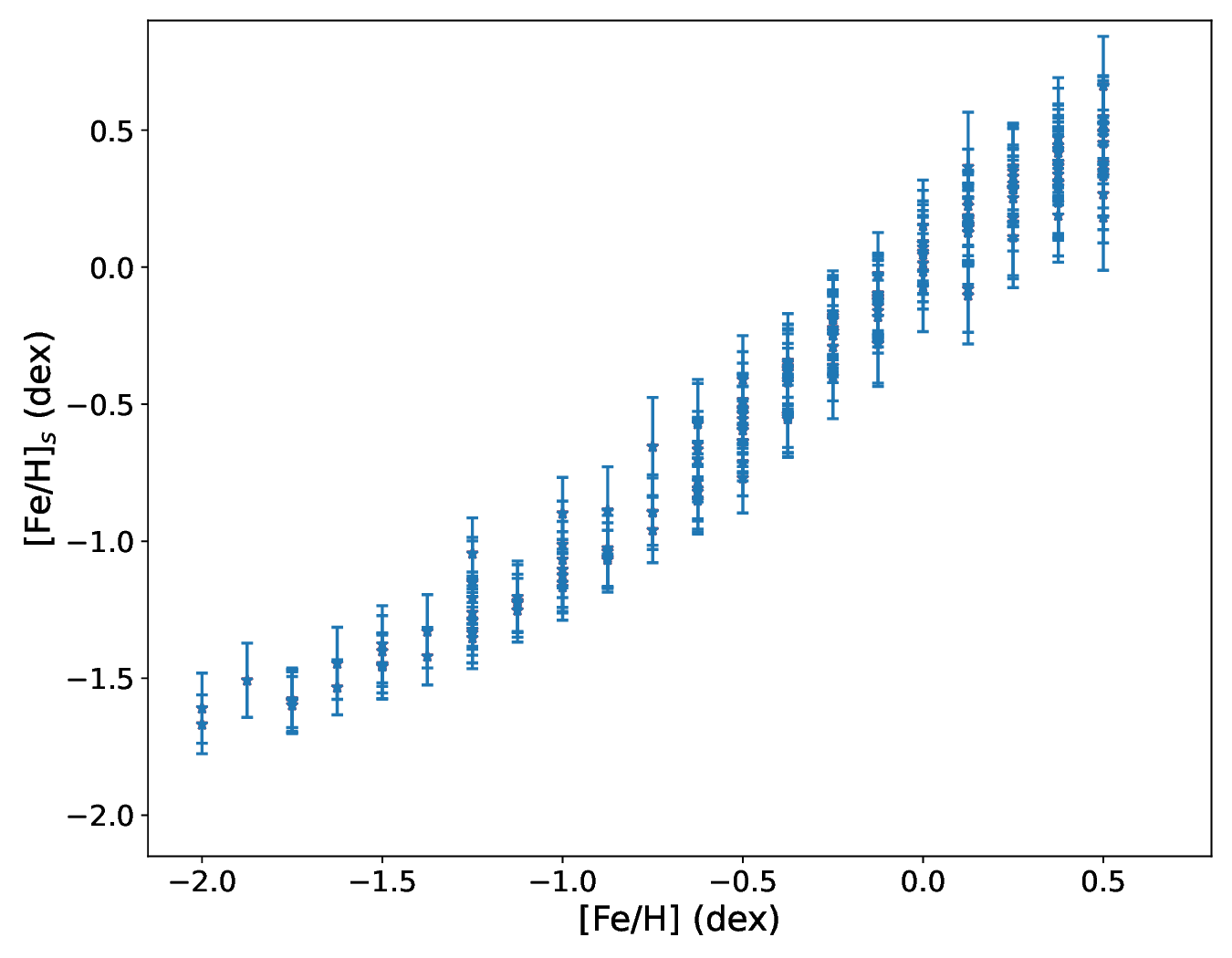}\hfill
    \caption{The scaling metallicity $\rm [Fe/H]_s$ (Equation \ref{eq.8}) versus the metallicity [Fe/H] for 176 HADS stars \citep{Netzel2022}.}\label{Figure 11}
\end{figure}

Stellar metallicity is the abundance of elements heavier than hydrogen and helium. However, in astrophysics, it is often written as the ratio of iron to hydrogen relative to the solar value ([Fe/H]; \cite{Kotoneva2002, Chruslinska2019}).
\citet{Netzel2022} obtained the [Fe/H] for 176 HADS stars by applying the MESA, Warsaw code, and optimization procedure.
These 176 HADS stars are mostly multimode ($\rm P0$, $\rm P1$, $\rm P2$,...) cases. The metallicity distribution for these 176 HADS stars ranges from -0.2 to 0.5 dex, which may be slightly related to the dispersion of period ratio ($\rm P1/\rm P0$ and $\rm P2/\rm P0$). \citet{Lv_2023} obtained a linear relation for period ratios ($\rm P1$/$\rm P0$ and $\rm P2$/$\rm P0$) and fundamental period $\rm P0$ (Figures 10 and 11 therein). Therefore, we may consider the $\rm P1$ and $\rm P2$ as a function of $\rm P0$ that allows us to investigate the scaling relation between metallicity, physical parameters, and fundamental period ($\rm P0$).

Using the mass ($M$), luminosity ($L$), fundamental period ($\rm P0$), and the effective temperature ($T_{\rm eff}$) of these 176 HADS stars, we determined the following scaling relation for [Fe/H] as 

\begin{equation}\label{eq.8}
    {{\rm [Fe/H]}_s = \log _{10}\left(\left (\frac {M}{M_\odot}\right)^{7.95\pm 0.15} \left (\frac {L}{L_\odot}\right)^{-1.83\pm 0.11} \left (\frac {\rm P0 (\rm day)}{0.064}\right)^{0.79\pm 0.14} \left (\frac {T_{\rm eff}}{K}\right)^{0.047\pm 0.02}\right)},
\end{equation}

where $\rm [Fe/H]_s$ is the scaling metallicity obtained by an optimization procedure for 176 HADS stars. The plus/minus values indicate the standard error of the power indexes in the model that obtained by the optimization algorithm.
Figure \ref{Figure 11} shows the comparison of [Fe/H]  and scaling $\rm [Fe/H]_s$ (Equation \ref{eq.8}) for HADS stars. Using Equation (\ref{eq.8}), we obtained the $\rm [Fe/H]_s$ of our seven newly identified HADS stars. 
We estimated the metallicity ranging from -0.62 to 0.37 dex (Table \ref{tab: Table 9}) for these seven HADS stars. Equation (\ref{eq.8}) is the scaling relation for the metallicity of multimode HADS, while three of our HADS (TIC 374753270, TIC 710783, and TIC 187386415) are the single-modes stars. Therefore, estimating the metallicity of three targets using Equation (\ref{eq.8}) may contain some uncertainty.

The four HADS stars (TIC 374753270, TIC 148357344, TIC 160120432, and TIC 278119167) show lower iron abundance than the Sun, while the rest of the three HADS (TIC 130474019, TIC 710783, and TIC 187386415) demonstrate positive [Fe/H]. The metallicities of these seven HADS stars are more significant than -0.5 dex, indicating metal-rich behavior \citep{McNamara2011}. The metal-rich behavior is an essential characteristic of the evolved old stars.

\begin{table}
\begin{threeparttable}
        \centering
        \
        \caption{TIC number, mass ($M/M_{\odot}$), radius ($R/R_{\odot}$), surface gravity (log g), effective temperature ($T_{\rm eff}$), absolute $V$-band magnitude ($M_{\rm V}$), and metallicity ($\rm [Fe/H]_s$) for seven HADS stars. }\label{tab: Table 9}
       \begin{tabular}{|c|c|c|c|c|c|c|}  
       \hline
      TIC number & $\
      M/M_{\odot}$ & $R/ R_{\odot}$ & log g & $T_{\rm eff}$  &$M_{\rm V}$& $\rm [Fe/H]_s$ \\
      \hline
       & &  &  (cm/$\rm s^2$)&  (K) &&  (dex)\\
      \hline
     374753270 & $1.60\pm20$&$3.04\pm 0.22$& $3.67\pm0.11$ &$7166\pm304.2$ & $1.61\pm0.16$ &$-0.451\pm0.161$\\
     130474019 &$1.76\pm0.21$  &$1.93\pm0.09$ &$4.11\pm0.10$  & $7573\pm197.5$ &$2.20\pm0.18$ &$0.243\pm0.135$\\
     148357344 &$1.61\pm0.21$& $2.99\pm0.22$ &$3.69\pm0.06$&$7183\pm270$&$1.51\pm0.23$ &$-0.465\pm0.160$\\ 
     160120432 & $1.59\pm0.19$ & $2.51\pm0.13$ & $3.84\pm0.08$&$7138\pm153.8$ &$1.61\pm0.16$ &$-0.266\pm0.143$\\ 
     278119167 &$1.68\pm0.2$ &$2.68\pm0.13$& $3.80\pm0.08$&$7369\pm185.1$ &$1.05\pm0.16$ &$-0.271\pm0.154$\\ 
     710783 &$1.77\pm0.23$  &$1.92\pm0.09$  & $4.11\pm0.07$&$7604\pm150.9$  &$2.04\pm0.22$ &$0.237\pm0.135$\\  
     187386415 &$1.67\pm0.20$ & $1.93\pm0.16$ &$4.08\pm0.08$ &$7340\pm97.2$ &$2.20\pm0.08$&$0.098\pm130$\\ 
    \hline
  \end{tabular}
  \begin{tablenotes}
	\item{ Note.These physical parameters (mass, radius, surface gravity, and effective temperature) of seven targets are given by \citet{Stassun2018}. Using Equations (\ref{eq.3}, \ref{eq.8}),  the absolute $V$-band magnitude ($M_{\rm V}$) and metallicity ($\rm [Fe/H]_s$) were calculated.}
	
    \end{tablenotes}
\end{threeparttable}
\end{table}

\section{conclusion}\label{sec:conclusion}
To probe the new HADS stars from recent \tess\ observations of \dsct\ stars, we recognized seven new targets, namely TIC 374753270, TIC 13047401, TIC 148357344, TIC 160120432, TIC 278119167, TIC 71083, and TIC 187386415, which show the peak-to-peak amplitude slightly more significant than 0.3 mag in their light curves. Using Equation (\ref{eq.3}) and the \Gaia\ DR3 parallaxes, the $V$-band absolute magnitude was calculated for these seven stars. Applying the \tess\ effective temperatures from \cite{Stassun2018} and the obtained absolute magnitudes, these seven HADS stars were located in the H-R diagram. They lie in the \dsct\ stars' stability strip of the H-R diagram (Figure \ref{Figure 2}) close to the cooler boundary.

The HADS stars were expected to show a simple amplitude spectrum, including one to a few large amplitudes. The period ratios, {$Q$} values, and PL relation between the fundamental radial frequency and luminosity were used to show the validity of mode information for these seven stars.
We applied the frequency analysis to obtain information on the significant frequencies of these stars. The significant frequencies followed an S/N greater than 5.2. Our analyses show that the HADS stars were categorized into single modes (TIC 374753270, TIC 710783, and TIC 187386415), double modes (TIC 130474019 and TIC 160120432), and triple modes (TIC 148357344 and TIC 278119167). We determined the variety of harmonic and combination frequencies in the amplitude spectrum of targets (Tables \ref{tab: Table 2}-\ref{tab: Table 8}) as well as nonradial oscillation frequencies for most of our targets. 
We derived a scaling relation using 176 HADS stars from \cite{Netzel2022} between metallicity ([Fe/H]) and physical parameters (mass ($M$), luminosity ($L$), and the effective temperature ($T_{\rm eff}$)) and fundamental period ($\rm P0$). Using Equation (\ref{eq.8}), we estimated the metallicity of the seven HADS stars. The metallicity of all seven stars is greater than -0.62 dex, as expected of the HADS stars.

\section{Acknowledgments}
This paper includes data collected by the \tess\ mission, which are publicly available from the Mikulski Archive for Space Telescopes (MAST). Funding for the \tess\ mission is provided by the NASA Explorer Program. Funding for the \tess\ Asteroseismic Science Operations Centre is provided by the Danish National Research Foundation (grant agreement
No.: DNRF106), ESA PRODEX (PEA 4000119301),
and Stellar Astrophysics Centre (SAC) at Aarhus University.
We would like to thank the \Gaia\ team for providing accurate data from the European Space Agency (ESA) mission \Gaia, processed by the \Gaia\ Data Processing and Analysis Consortium (DPAC). Funding
for the DPAC has been provided by national institutions, in particular, the institutions participating in the \Gaia\ Multilateral Agreement.
This work has been supported by the Iran National
Science Foundation (INSF) under grant No. 4002562.
E.Z. expresses gratitude for that. Also, E.Z. acknowledges financial support from project PID2019-107061GB-C63 from the ‘Programas Estatales de Generación de Conocimiento y Fortalecimiento Científico y Tecnológico del Sistema de I+D+i y de I+D+i Orientada a los Retos de la Sociedad’ and from the grant CEX2021-001131-S funded by MCIN/AEI/10.13039/501100011033.

%

\vspace{5mm}








\bibliography{references}{}
\bibliographystyle{aasjournal}



\end{document}